\documentclass[prb,aps,twocolumn,footinbib,superscriptaddress,floatfix,bibliography]{revtex4-1}

\usepackage{CJK}
\usepackage[colorlinks,bookmarks=false,citecolor=blue,linkcolor=red,urlcolor=blue]{hyperref}
\usepackage{color} 
\usepackage{graphicx}
\usepackage{float}
\usepackage{multirow}
\usepackage{amsmath}
\usepackage{bm}
\usepackage{amsfonts}
\usepackage{braket}
\usepackage{subfigure}
\usepackage{cleveref}

\DeclareMathOperator{\sech}{sech}

\begin{document}
\title{Spin ice in a general applied magnetic field: Kasteleyn transition, magnetic torque and rotational magnetocaloric effect}

\author{Mark Potts}
\affiliation{University of Oxford,
Dept. of Physics, 
Parks Road,
Oxford,
OX1 3PU}
\affiliation{Max Planck Institute for the Physics of Complex Systems, N{\"o}thnitzer Stra{\ss}e 38, Dresden 01187, Germany}

\author{Owen Benton}
\affiliation{Max Planck Institute for the Physics of Complex Systems, N{\"o}thnitzer Stra{\ss}e 38, Dresden 01187, Germany}

\begin{abstract}
Spin ice is a paradigmatic frustrated system
famous for the emergence
of magnetic monopoles and a large magnetic entropy at
low temperatures.
It exhibits unusual behavior in the presence of an external
magnetic field as a result of
the competition between the spin ice entropy and the Zeeman energy.
Studies of this have generally focused on fields applied along high symmetry directions: $[111]$, $[001]$, and $[110]$.
Here we consider a model of spin ice with external field in an arbitrary direction.
We find that the Kasteleyn transition known for $[001]$ fields,
appears also for general field directions and calculate the associated Kasteleyn temperature $T_K$ as a function of field direction.
$T_K$ is found to vanish, with a logarithmic
dependence on field angle, approaching certain lines of special
field directions.
We further investigate the thermodynamic properties of spin ice
for $T>T_K$, using a Husimi cactus approximation. 
As the system is cooled towards $T_K$ a large magnetic torque appears, tending to align the $[001]$ crystal axis with the external field.
The model also exhibits a rotational magnetocaloric effect: significant temperature changes can be obtained by adabiatically rotating the crystal relative to a fixed field.
\end{abstract}

\maketitle

\section{Introduction}
\label{sec:intro}

Spin ice exemplifies much of what is interesting
about frustrated systems \cite{harris97, Balents2010, henley10, castelnovo12}.
The co-existence of a large quasi-degeneracy
of ground states \cite{Ramirez1999, melko04, isakov05, giblin18} and strong correlations \cite{bramwell01-PRL87, isakov04, yavorskii08, Fennell2009, kanada09}
sets the stage for the emergence of exotic excitations: namely, magnetic monopoles \cite{Castelnovo2008, Morris2009, Jaubert2009, kadowaki09, ryzhkin13, kaiser18, dusad19, samarakoon22}.
The discovery of spin ice also served as an early
example of a theme which has grown in importance
in the years since: the interplay of frustration
and anisotropy \cite{fukazawa02}.

Magnetic anisotropy is the root of how spin ice can be frustrated, despite dominantly ferromagnetic interactions  \cite{bramwell98, moessner98-PRB57}. The importance of anisotropy is also seen through the diverse behaviors induced by applying a magnetic field 
along different crystal directions.
Fields along $\langle111\rangle$ induce an effectively two-dimensional, disordered, ``kagome ice'' state \cite{moessner03, hiroi03, sakakibara03, isakov04magcurve}; fields along $\langle110\rangle$ induce a division of the system into effectively one-dimensional chains \cite{hiroi03chains, clancy09,  guruciaga16, placke20};
while a field along $\langle100\rangle$ drives a Kasteleyn transition to an ordered state \cite{powell08, jaubert08, jaubert09-jpcs}.
Most studies of spin ice in an applied field have focused on fields oriented 
along \cite{hiroi03, sakakibara03, isakov04magcurve, hiroi03chains, clancy09,  guruciaga16, placke20, powell08, jaubert08, jaubert09-jpcs, harris98, pili21, baez16}, 
or close to \cite{moessner03, fennell07}, those high symmetry directions.
Here we give an account of the physics of an
idealized spin ice model, with a completely
general direction of external field.

We determine the ground state phase diagram as a function of applied field, showing that for non-fine-tuned choices of field direction there is a unique ground state with magnetisation along a $\langle 100 \rangle$ axis. 
At finite temperature, there is a Kasteleyn transition at $T=T_K$, separating the field induced order at $T<T_K$ from a Coulomb phase
at $T>T_K$. We determine the dependence of $T_K$
on the field direction, showing that it approaches zero in a singular fashion near the boundaries of the ground state phase diagram.

We then go on to study the thermodynamics of the system at $T>T_K$ using a Husimi tree approximation. Our account of the thermodynamics is focussed on the effects of applying an external field which is not aligned with a high symmetry direction of the crystal. We find that a large magnetic torque develops as $T_K$ is approached from above, as the exchange energy forces the system to align the magnetisation closer to a $\langle100\rangle$ axis and away from the magnetic field. Relatedly, we find a rotational magnetocaloric effect, in which large changes in temperature can be driven by adiabatic rotation of the crystal relative to the field.

The conventional magnetocaloric effect (MCE) -- change in temperature driven by a change in 
applied field strength -- has long been known as a useful probe  of frustrated magnetic systems \cite{zhitomirsky03, kohama12, tokiwa13, manni14}, including spin ice \cite{aoki04, orendac07}, and as a potential basis for cleaner refrigeration technology \cite{pecharsky97, gschneidner08, balli17}.
By contrast, the rotational magnetocaloric effect (RMCE) has begun to attract significant attention only relatively recently.
The RMCE could present certain technological
advantages over the conventional MCE \cite{balli14, balli17} and recent 
developments in measurement techniques make it
increasingly practical to use RMCE as a probe of novel physics in anisotropic frustrated systems \cite{kittaka18, kittaka21}.
Here we will show that an idealized model of
spin ice predicts an appreciable RMCE, and that 
interactions enhance the RMCE by an order of magnitude above what would be expected from a simple paramagnet with the same symmetry and single-ion anisotropy.

The Article is organized as follows: in Section
\ref{sec:ground_states} we introduce the model and determine the ground state phase diagram; in Section \ref{sec:kasteleyn} we calculate the Kasteleyn temperature. $T_K$, as a function of field direction; in Section \ref{sec:thermodynamics}
we discuss the thermodynamics at $T>T_K$ including the magnetic torque and RMCE; 
before concluding in Section \ref{sec:summary}.

\section{Model and Ground states}
\label{sec:ground_states}

We consider a nearest-neighbor model
for spin ice
\begin{equation}
    H=-J\sum_{\left\langle{i,j}\right\rangle}\vec{S}_i\cdot\vec{S}_j-\vec{h} \cdot \sum_{i}\vec{S}_i
    - D_{\rm SI} \sum_i \left( \vec{S_i} \cdot \vec{e}_i \right)^2
    \label{eq: Hamiltonian}
\end{equation}
where the first term is a ferromagnetic nearest neighbor exchange interaction, the second term is the
Zeeman energy and the third term is an easy-axis single-ion anisotropy. 
Throughout this Article, we
take $D_{SI}, J >0$ and
consider the limit $D_{SI}\gg J \gg h$.

\begin{figure}
    \centering
    \includegraphics[width=6cm]{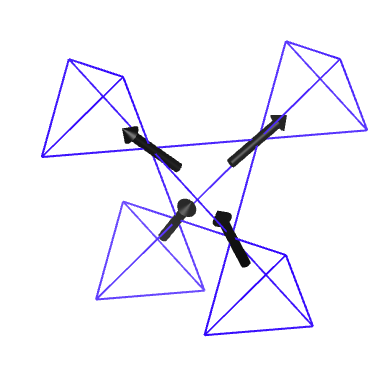}
    \caption{The 2-in-2-out `ice-rule' obeyed by spin-ice on the pyrochlore lattice. All tetrahedra must obey this rule as long as the energy scale associated with the ferromagnetic exchange interaction $J$ is much larger than $T$ and $h$.}
    \label{fig: spin ice 2-in-2-out}
\end{figure}

The strong easy-axis anisotropy $D_{SI}\gg J$ aligns the direction of the classical spin $\vec{S}_i$ with the line connecting the centers of the two tetrahedra
sharing site $i$ [Fig. \ref{fig: spin ice 2-in-2-out}].
The four spins in a unit cell have different easy-axis
directions $\vec{e}_i$. Numbering the sites in a unit cell from 1 to 4, we choose coordinates where:
\begin{eqnarray}
    \vec{e}_1=
    \begin{pmatrix}
    \frac{1}{\sqrt{3}} \\
    \frac{1}{\sqrt{3}} \\
    \frac{1}{\sqrt{3}} \\
    \end{pmatrix}, \ \vec{e}_2=
    \begin{pmatrix}
    \frac{1}{\sqrt{3}} \\
    -\frac{1}{\sqrt{3}} \\
    -\frac{1}{\sqrt{3}} \\
    \end{pmatrix}, \ \nonumber \\
    \vec{e}_3=
    \begin{pmatrix}
    -\frac{1}{\sqrt{3}} \\
    \frac{1}{\sqrt{3}} \\
    -\frac{1}{\sqrt{3}} \\
    \end{pmatrix}, \ \vec{e}_4=
    \begin{pmatrix}
    -\frac{1}{\sqrt{3}} \\
    -\frac{1}{\sqrt{3}}\\
    \frac{1}{\sqrt{3}} \\
    \end{pmatrix}
    \label{eq: spin directions}
\end{eqnarray}
Each spin thus has two orientations, which we can parameterise using an Ising variable $\sigma_i$:
\begin{equation}
    \vec{S}_i=\sigma_i \vec{e}_i
    \label{eq:sigma_introduction}
\end{equation}

Using that $\vec{e}_i \cdot \vec{e}_j = -\frac{1}{3}$ for neighboring sites $i,j$, and dropping an unimportant constant term, the Hamiltonian becomes:
\begin{equation}
    H=\frac{J}{6}\sum_{\Delta}\left(\sum_{i\in\Delta}\sigma_i\right)^2-\sum_{i}\sigma_i \left(\vec{h} \cdot \vec{e}_i \right) \label{eq: Hamiltonian /w explicit frustration}
\end{equation}

Here $\Delta$ indexes tetrahedra in the pyrochlore lattice. We see that for $J \gg h$, the ground state must be one in which $\sigma_i$ sums to zero on each tetrahedron or, equivalently, where each tetrahedron has two spins pointing in, and two pointing out. 
This is the ``ice rule'', illustrated in Fig. \ref{fig: spin ice 2-in-2-out}.
We will work in a limit where $J\rightarrow\infty$, and the ice rule will be absolutely obeyed.

The lattice structure in Fig. \ref{fig: spin ice 2-in-2-out} shows tetrahedra of two orientations. We
will refer to these as `A' and `B' tetrahedra.
The Zeeman term of the Hamiltonian can be expressed 
as a sum over `A' tetrahedra:
\begin{eqnarray}
   && H=\sum_{\Delta \in A }
    \left[ \frac{J}{6}\left(\sum_{i\in\Delta}\sigma_i\right)^2-\sum_{i\in\Delta} \sigma_i \left(\vec{h} \cdot \vec{e}_i \right)\right] + \nonumber \\
  && \qquad \qquad  \sum_{\Delta \in B }
 \frac{J}{6}\left(\sum_{i\in\Delta}\sigma_i\right)^2 
    \label{eq: Hamiltonian /w Zeeman summed over A}
\end{eqnarray}

Parametrising the direction of the external field with
angles $\theta$ and $\varphi$:
\begin{eqnarray}
\vec{h}=\begin{pmatrix}
\cos(\varphi) \sin(\theta) \\
\sin(\varphi) \sin(\theta) \\
\cos(\theta)
\end{pmatrix}
\label{eq:h_direction}
\end{eqnarray}
and using Eq. (\ref{eq: spin directions}), we have  that for one `A' tetrahedron, the Zeeman energy is:
\begin{multline}
    H_{Z}=-\frac{h}{\sqrt{3}}\sigma_1\left[\sin(\theta)\cos(\varphi)+\sin(\theta)\sin(\varphi)+\cos(\theta)\right] \\
    -\frac{h}{\sqrt{3}}\sigma_2\left[\sin(\theta)\cos(\varphi)-\sin(\theta)\sin(\varphi)-\cos(\theta)\right] \\ -\frac{h}{\sqrt{3}}\sigma_3\left[-\sin(\theta)\cos(\varphi)+\sin(\theta)\sin(\varphi)-\cos(\theta)\right] \\
    -\frac{h}{\sqrt{3}}\sigma_4\left[-\sin(\theta)\cos(\varphi)-\sin(\theta)\sin(\varphi)+\cos(\theta)\right]
\label{eq: Zeeman Hamiltonian full}
\end{multline}

We can then find  the ground state spin configuration as a function of $\theta$ and $\varphi$
for a single $A$ tetrahedron.
Where this single tetrahedron ground state is non-degenerate, the ground state of the whole lattice
is then found simply by tiling the single tetrahedron
ground state over all `A' tetrahedra. 
One only needs to check that this tiling does not
induce a violation of the ice rule on the `B' tetrahedra, but this can readily be verified.

Where the ground state of the `A' tetrahedra is
degenerate, there may be many ways to tile
the single tetrahedron ground states across the lattice, while maintaining consistency with the ice rule on the `B' tetrahedra.

\begin{figure}[h]
\centering
    \subfigure[]{
    \includegraphics[width=7cm]{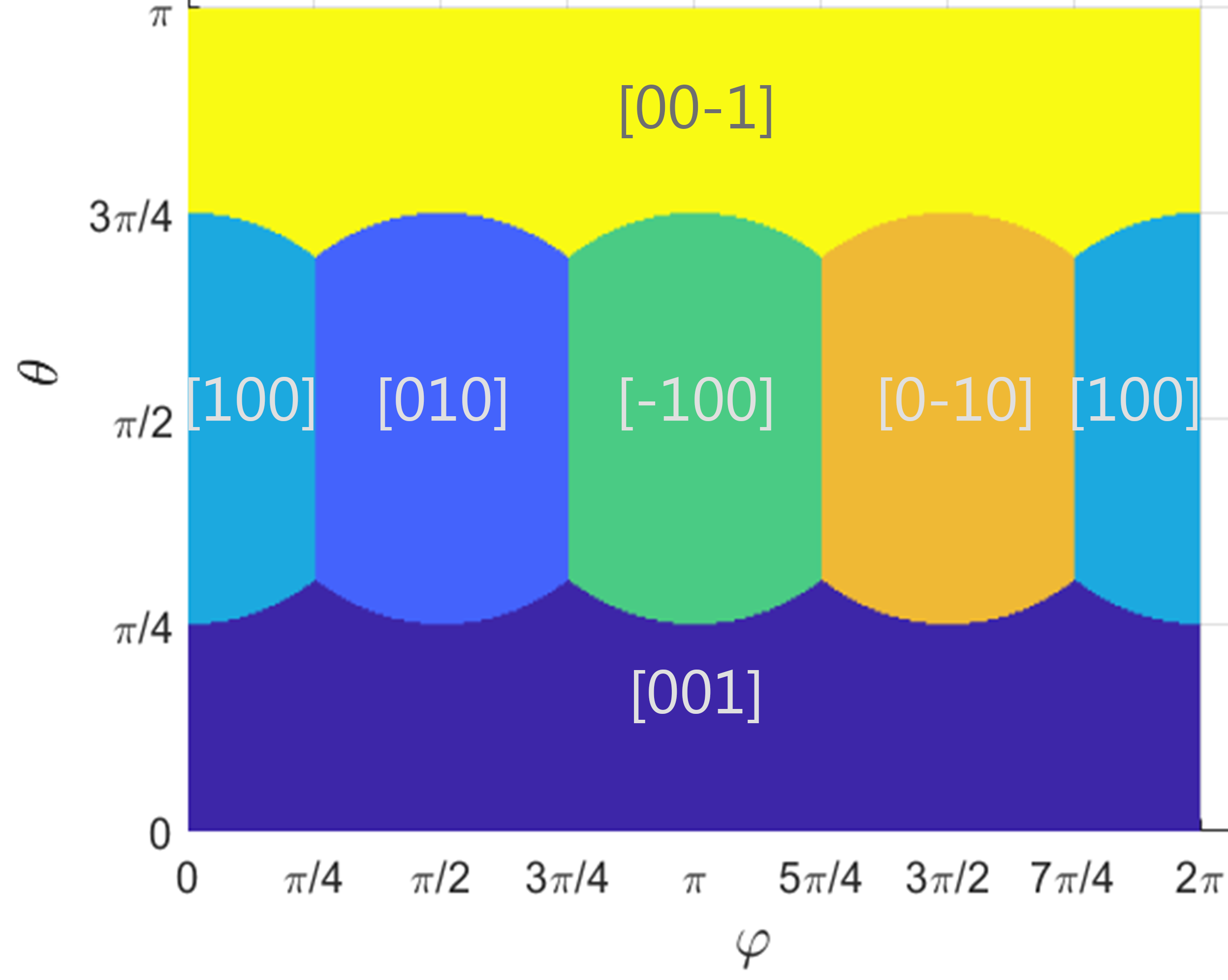}}\\
    \subfigure[]{
    \includegraphics[width=7cm]{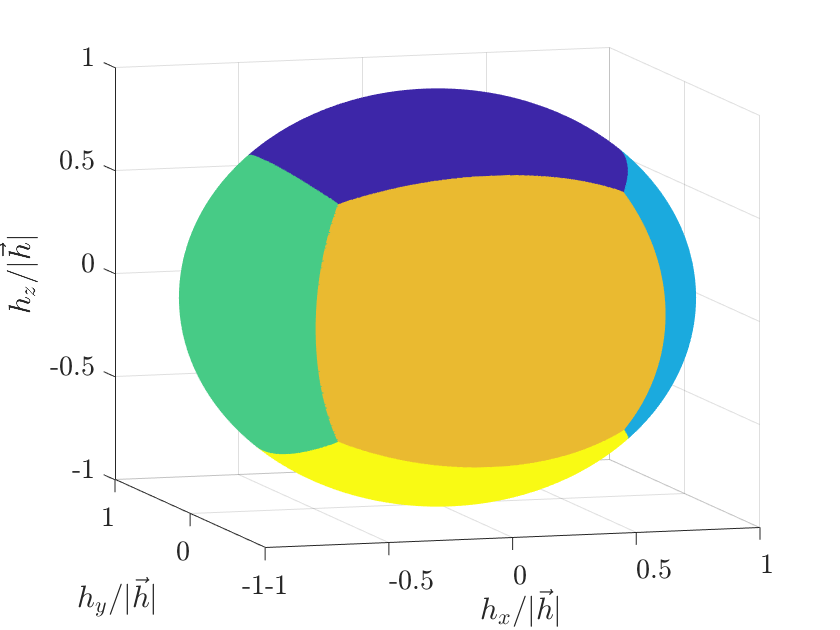}}
    \caption{Magnetic ground state phase diagram as a function of external field direction. (a) Phase diagram in the $\theta$-$\varphi$ plane [Eq. (\ref{eq:h_direction})]. (b) Phase diagram mapped onto the unit sphere. Each phase is labelled with the direction of the magnetisation in the ground state. The spin configurations in each phase are simply related to one another by rotation of the entire system, and the full physics can be investigated by considering just one of these phases and its boundaries.
    }
    \label{fig:Magnetic phase diagram}
\end{figure}

A phase diagram, mapping out the ground states for general field directions with $h \ll J$ is shown in
Fig. \ref{fig:Magnetic phase diagram}.
There are 6 distinct phases which occupy a finite 
area of the phase diagram, which are labelled by their magnetisation direction.
They correspond to uniform tilings across the lattice of each of the six possible single-tetrahedron ice-rule states.

The phase diagram contains lines along which two
single tetrahedron ground states are degenerate.
Along these lines the system splits into two sets of
independent 1D chains $\alpha$ and $\beta$.
The configuration on the $\alpha$ chains is fixed by
the applied field, whereas each $\beta$ chain has
an independent two-fold degeneracy.
This is well known for the case of a $\langle 110 \rangle$ field \cite{hiroi03chains, clancy09, guruciaga16, placke20}.
It is interesting to note that the chain degeneracy
actually persists along a line of field directions including,
but not limited to, the $\langle 110 \rangle$ case.

The points on the phase diagram where three single tetrahedron ground states become degenerate correspond
to $\langle 111 \rangle$ fields, and the well studied
case of kagome ice \cite{hiroi03, udagawa02, moessner03}.
In this case, the system splits into independent kagome
planes, and their remains an extensive residual entropy.

For the remainder of this paper we will principally consider the generic case, where the single tetrahedron ground state is non-degenerate, although we will also note the behavior approaching the degenerate limits.

\section{Kasteleyn Transition Temperature}
\label{sec:kasteleyn}

In this section we consider the finite-temperature phase transition between the field induced ordered phase and the Coulomb phase.
This transition is a Kasteleyn transition, and has been
studied previously for the case of spin ice in a $\langle001\rangle$ field
\cite{powell08, jaubert08}, 
and fields close to the $\langle111\rangle$ direction \cite{moessner03}.
Here we give a generalisation of this to field directions not aligned with a high symmetry direction of the crystal.

We consider field directions such that  
the largest of the three components of $\vec{h}$ is the
$z$-component.
In this case,
the ground state has magnetisation along the $[001]$ direction [see Fig. \ref{fig:Magnetic phase diagram}].
Results for other directions can be obtained straightforwardly by applying cubic rotations to this case.

\begin{figure}
    \centering
    \includegraphics[width=6cm]{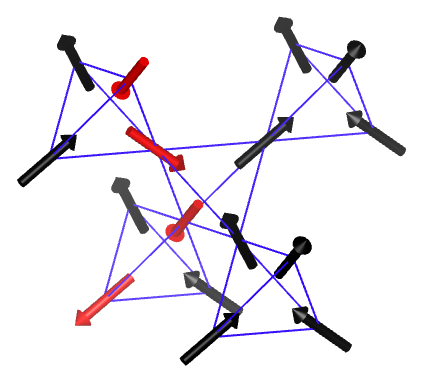}
    \caption{A spin configuration containing a string excitation. The string (highlighted in red) spans the entire system, and costs energy proportional to the linear system size.}
    \label{fig:Spin ice string}
\end{figure}

The ground state is a configuration in which every  tetrahedron is in the same 2-in-2-out state with magnetisation along $[001]$.
Because we take $J \to \infty$ and do not allow violations of the ice rule, excitations are not single spin flips, but extended strings of flipped spins, spanning the entire system [Fig. \ref{fig:Spin ice string}].
If ice rule violating tetrahedra are allowed, the sharp Kasteleyn transition discussed below becomes a crossover.

The energy cost of a string excitation is proportional to its length, because every flipped spin increases the Zeeman energy. Therefore, at sufficiently low temperature, string excitations are completely suppressed in the thermodynamic limit.

However, the entropy of the string is also proportional to its length, because at every successive layer through the system the string can go one of two ways. The total free energy therefore
has competing contributions, and at some finite temperature $T_K$, the sign of the free energy per unit length of string changes. For $T>T_K$, introducing strings decreases the free energy and strings therefore proliferate, destroying the ordered state. This is the 
Kasteleyn transition.

For a general magnetic field direction, where the Zeeman energy is not the same for all sites, the energy of the string depends on the path it takes through the system.
This is different to the case of the $[001]$ field where all string paths have the same energy
per unit length \cite{jaubert08, powell08}.
This has to be taken into account when constructing the string free energy.

\begin{figure}
        \centering
        \includegraphics[width=8cm]{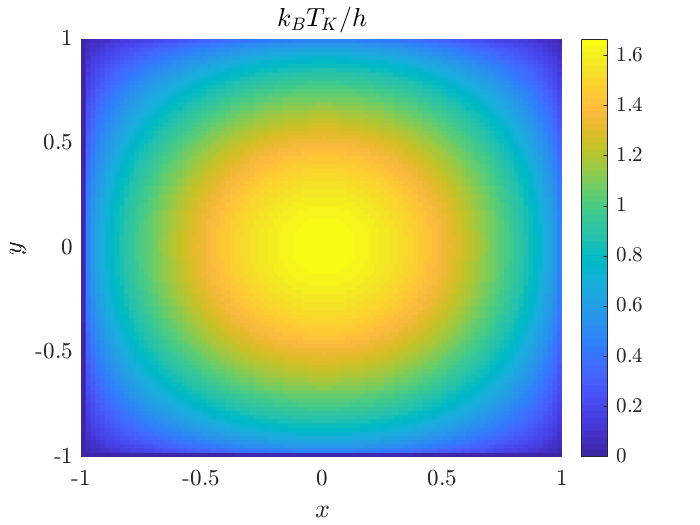}
        \caption{Variation of the Kasteleyn temperature as a function of field direction. The Kasteleyn temperature is given in units of $h/k_B$. The centre of the plot corresponds to the [001] field direction, where the result $T_K=\frac{2h}{\sqrt{3}\log{2}}$ is reproduced\cite{powell08}. The mapping between the x-y coordinates and $\theta$ and $\varphi$ is given in Eqs. (\ref{eq: x parameterisation})-(\ref{eq: theta parameterisation}).
        $T_K$ approaches zero at the edge of the plot, which corresponds to the phase boundaries of Fig. \ref{fig:Magnetic phase diagram}.
        }
        \label{fig:Kasteleyn_Temp_plot}
\end{figure}
\begin{figure*}
        \centering
        \subfigure[  ]{
        \includegraphics[width=8cm]{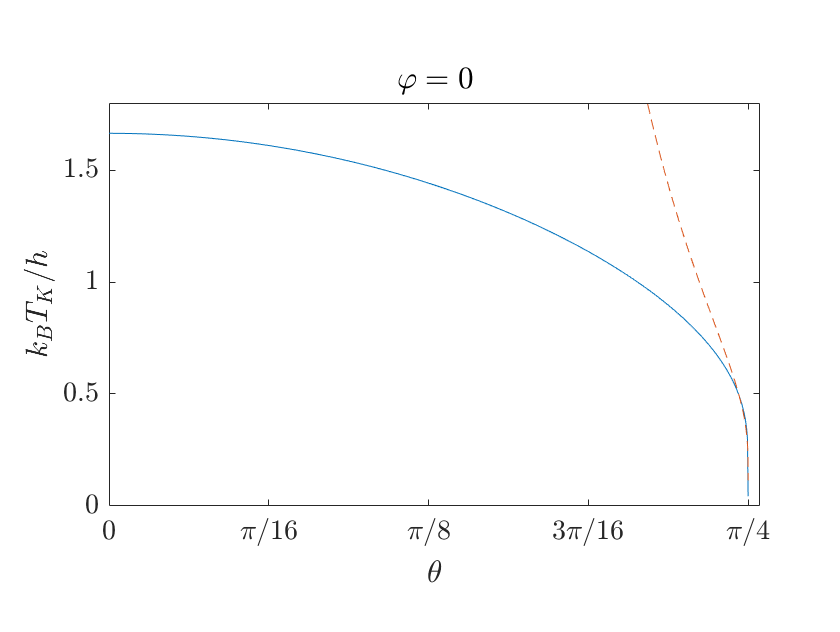}}
        \subfigure[ ]{
        \includegraphics[width=8cm]{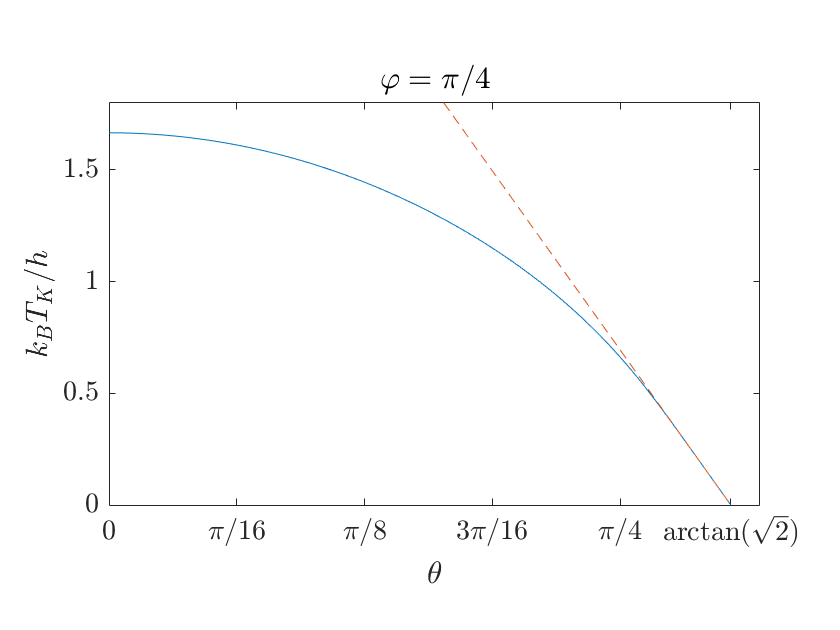}}
        \caption{Variation of the Kasteleyn transition temperature as a function of field angle $\theta$ for   $\varphi=0$ [(a)]
        and $\varphi=\frac{\pi}{4}$
        [(b)]. 
        The dashed lines show the asymptotic behavior of $T_K$ approaching the magnetic phase boundaries of Fig. \ref{fig:Magnetic phase diagram}, as predicted by Eq. (\ref{eq:pb_log}) for $\varphi=0$ and Eq. (\ref{eq:pb_linear}) for $\varphi=\frac{\pi}{4}$.
        $T_K$ vanishes as $\frac{-1}{\log(\delta)}$ approaching the phase boundaries for generic values of $\varphi$, including $\varphi=0$, where 
        $\delta$ is the distance from the phase boundary.
        For the special case $\varphi=\frac{\pi}{4}$, $T_K$ vanishes linearly towards the kagome ice point\cite{moessner03} at $\theta=\arctan(\sqrt{2})$.
        }
        \label{fig:T_k_slices}
\end{figure*}

In general, each of the four sites in the pyrochlore unit cell has a different Zeeman energy.
Dividing the system into layers normal to the $[001]$
direction, sites with spins in two of these sublattices share one layer, and sites with spins in the remaining two sublattices share the next.

We construct the partition function for a single two layers at a time.
Considering all the paths a string could take through two consecutive layers,  the two layer partition function is:
\begin{multline}
    Z_{2}=2\exp\left(-\frac{-4h\beta}{\sqrt{3}}\cos{\theta}\right)\left\lbrace\cosh\left(\frac{4h\beta}{\sqrt{3}}\sin(\theta)\cos(\varphi)\right) \right.\\ \left. +\cosh\left(\frac{4h\beta}{\sqrt{3}}\sin(\theta)\sin(\varphi)\right)\right\rbrace
    \label{eq: partition function}
\end{multline}

From this we can calculate the free energy of a string excitation per two segments:
\begin{multline}
    F=\frac{4h}{\sqrt{3}}\cos{\theta}-\frac{1}{\beta}\log(2) \\
-\frac{1}{\beta}\log\left(\cosh\left(\frac{4h\beta}{\sqrt{3}}\sin(\theta)\cos(\varphi)\right) \right. \\ \left.+\cosh\left(\frac{4h\beta}{\sqrt{3}}\sin(\theta)\sin(\varphi)\right)\right)
    \label{eq: String free energy}
\end{multline}

Setting $F=0$ we find the following expression \mbox{for $\beta_K = \frac{1}{k_B T_K}$}
\begin{multline}
    \frac{4h}{\sqrt{3}}\cos{\theta}-\frac{1}{\beta_K}\log(2) \\
-\frac{1}{\beta_K}\log\left(\cosh\left(\frac{4h\beta_K}{\sqrt{3}}\sin(\theta)\cos(\varphi)\right) \right. \\ \left.+\cosh\left(\frac{4h\beta_K}{\sqrt{3}}\sin(\theta)\sin(\varphi)\right)\right)=0.
    \label{eq: beta_K}
\end{multline}

Taking the limit $\theta=0$, Eq. (\ref{eq: beta_K})
reproduces the known result
\cite{powell08} for the case of $\vec{h} \parallel [0,0,1]$:
\begin{eqnarray}
k_B T_K (\theta=0) = 
\frac{2 h}{\sqrt{3}\log(2)}.
\end{eqnarray}

For more general field directions, Eq. (\ref{eq: beta_K}) can be solved numerically to
obtain the dependence of $T_K$ on the applied field
direction. This is shown in Figs. \ref{fig:Kasteleyn_Temp_plot} and \ref{fig:T_k_slices}.
In these figures we make use of the following parameterisation for the angles $\theta$ and $\varphi$, understood as projecting points on the unit sphere onto one face of a unit cube circumscribing it.

\begin{align}
    x=&\tan(\theta)\cos(\varphi) \label{eq: x parameterisation}\\
    y=&\tan(\theta)\sin(\varphi) \label{eq: y parameterisation}\\
    \tan(\varphi)=&\frac{y}{x} \label{eq: phi parameteristaion}\\
    \cos(\theta)=&\frac{1}{\sqrt{x^2+y^2+1}} \label{eq: theta parameterisation}
\end{align}

This mapping transforms the magnetic phase boundaries of Fig. \ref{fig:Magnetic phase diagram} onto the edges of the cube. 

$T_K$ actually vanishes at these phase boundaries.
This can be seen by taking the limit $T_K \ll h$ in Eq. (\ref{eq: beta_K}):
\begin{eqnarray}
&&\frac{4h}{\sqrt{3}} \cos\theta
- k_B T_K \log \bigg\{ 
\bigg[
\exp\bigg(\frac{4h}{\sqrt{3}k_B T_K} \sin(\theta) |\cos(\varphi)|\bigg)
+ \nonumber \\
&&\exp\bigg(\frac{4h}{\sqrt{3}k_B T_K} \sin(\theta) |\sin(\varphi)|
\bigg)
\bigg]
\bigg\}=0
\label{eq:TK1}
\end{eqnarray}

Considering first the case $|\cos(\varphi)|>|\sin(\varphi)|$,
the logarithm can be expanded to obtain for $T_K \ll h$ to obtain:
\begin{eqnarray}
&&\frac{4h}{\sqrt{3}} ( \cos\theta - \sin(\theta) |\cos(\varphi)|
)
\nonumber \\
&& - k_B T_K 
\exp\bigg(-\frac{4h}{\sqrt{3}k_B T_K}
\sin(\theta)\big( |\cos(\varphi)| - |\sin(\varphi)|\big)\bigg) 
\nonumber \\
&&=0
\label{eq:TK2}
\end{eqnarray}
from which we can see that $T_K$ vanishes when
\begin{eqnarray}
\cos(\theta) = \sin(\theta) |\cos(\varphi)|
\label{eq:pb1}.
\end{eqnarray}

Detuning $\theta$ away from this phase boundary gives by an amount $\delta$, gives
rise to a logarithmic behavior $T_K \sim \frac{1}{\log\left( \frac{1}{\delta} \right)}$.
To see this, we define $\theta_0 (\varphi)$ to be the value
of $\theta$ that satisfies Eq. (\ref{eq:pb1})
and write $\theta=\theta_0 - \delta $.
Expanding Eq. \ref{eq:TK2} for small $\delta$, still with $T_K \ll h$
\begin{eqnarray}
&&\frac{\delta}{\sqrt{3}} \bigg[
 \cos\big(\theta_0 (\varphi)\big) |\cos(\varphi)|
 -\sin\big(\theta_0 (\varphi)\big) 
\bigg]
= \nonumber \\
&&
 \frac{k_B T_K}{h} \exp\bigg(-\frac{4h}{\sqrt{3}k_B T_K}
\sin(\theta_0 (\varphi))( |\cos(\varphi)| - |\sin(\varphi)|)\bigg) 
\nonumber \\
&& \\
&&\implies 
k_B T_K \approx
\frac{4h\sin(\theta_0 (\varphi))( |\cos(\varphi)| - |\sin(\varphi)|))  }{ \sqrt{3} \log\left(\frac{1}{\delta} \right) }.
\label{eq:pb_log}
\end{eqnarray}
This asymptotic result is compared with the numerical solution to Eq. (\ref{eq: beta_K})
in Fig. \ref{fig:T_k_slices}(a).

A similar result is obtained for the case 
$|\cos(\varphi)|<|\sin(\varphi)|$,
with $T_K$ vanishing when
\begin{eqnarray}
\cos(\theta) = \sin(\theta) |\sin(\varphi)|
\label{eq:pb2}.
\end{eqnarray}
and depending logarithmically on the variation in
$\theta$ away from the phase boundary.

A qualitatively different behavior of $T_K$ is obtained
for the special case $|\cos(\varphi)|=|\sin(\varphi)|$.
In this case, varying $\theta$ tunes the system towards the
``kagome ice'' point at $\theta = \arctan\left(\sqrt{2}\right)$ and
$T_K$ vanishes linearly as $\theta$ approaches this limit.
To see this we set $\varphi=\frac{\pi}{4}$ in Eq.
(\ref{eq: beta_K}) and obtain:
\begin{eqnarray}
\frac{4h}{\sqrt{6}} ( \sqrt{2}\cos(\theta) - \sin(\theta)  )= 
k_B T_K \log(2).
\end{eqnarray}
Setting $\theta = \arctan\left(\sqrt{2}\right) - \delta$
and expanding for small $\delta$ then gives:
\begin{eqnarray}
k_B T_K=
\frac{2 \sqrt{2} h \delta }{ \log(2)}
\label{eq:pb_linear}
\end{eqnarray}
in agreement with the result in Ref. \onlinecite{moessner03} for fields
close to a $\langle111\rangle$ axis.
$T_K$ thus vanishes linearly approaching the kagome ice
point $\theta=\arctan(\sqrt{2}), \varphi=\frac{\pi}{4}$.
This asymptotic result is shown 
in Fig. \ref{fig:T_k_slices}(b).

Having now determined the behavior of $T_K$ as the field direction is varied, we will turn to consider the
thermodynamics of spin ice as $T_K$ is approached from above, for generic applied field directions.

\section{Thermodynamics above the Kasteleyn transition}
\label{sec:thermodynamics}

In this Section we study the thermodynamics of the Coulomb phase
as the system is cooled towards $T_K$.
To do this, we make use of the Husimi tree approximation
\cite{jaubert08, jaubert-thesis, jaubert14, otsuka18,jurcisinova17}, which
is described in Section \ref{subsec:husimi}; before presenting results for the heat capacity and entropy [Section \ref{subsec:heatcapacity}],
magnetisation and magnetic torque [Section \ref{subsec:magnetisation}]
and the rotational magnetocaloric effect [Section \ref{subsec:magnetocaloric}].

\subsection{Husimi Tree Approach}
\label{subsec:husimi}

\begin{figure}
    \centering
    \includegraphics[width=6cm]{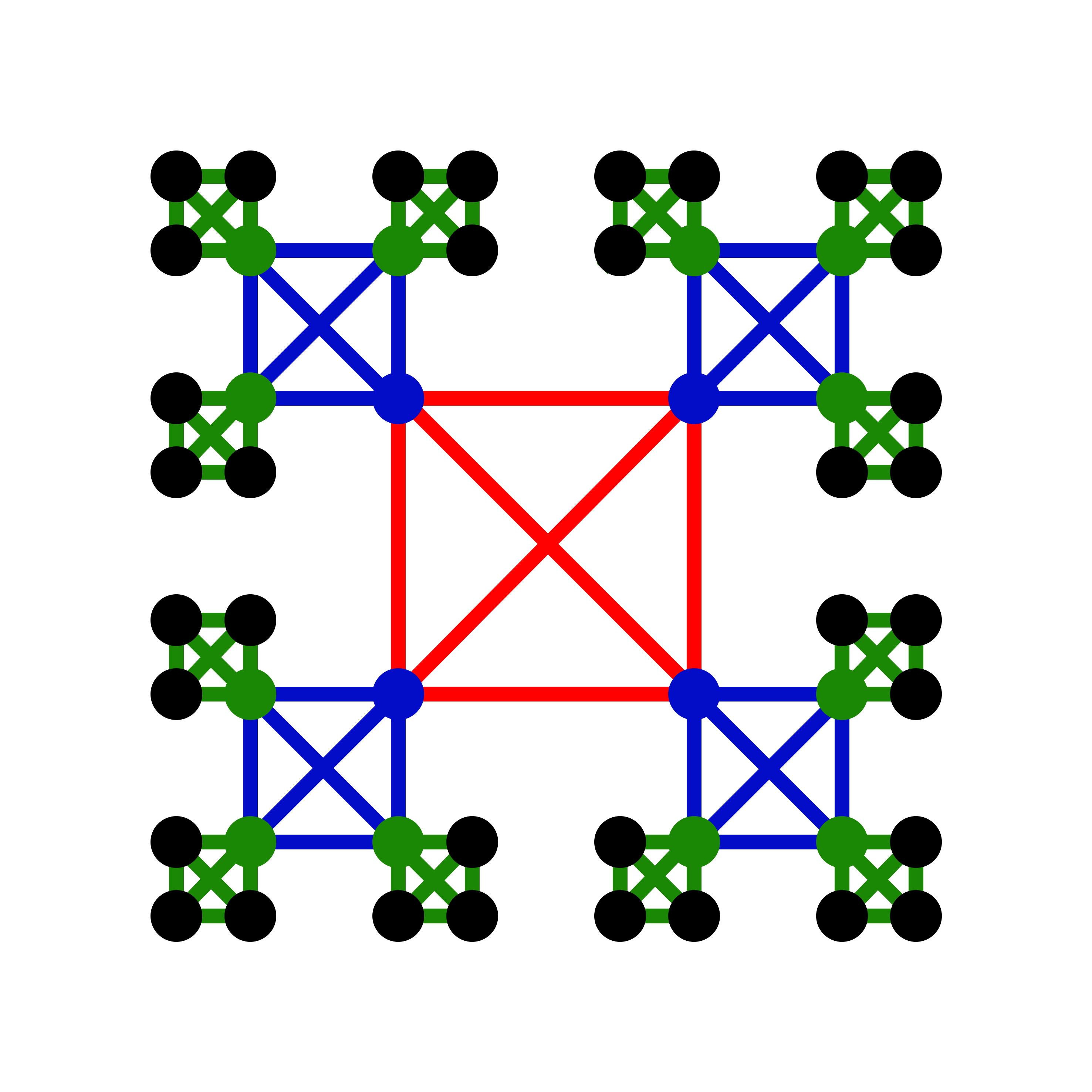}
    \caption{Diagram of the Husimi tree construction for three shells. The dots represent spins, and the boxes tetrahedra. Tetrahedra are a shown in a two-dimensional projection to highlight the tree's topology. The outermost, zeroth shell of spins, is shown in black, the first in green, and the second in blue.}
    \label{fig:Husimi tree}
\end{figure}

The Husimi tree approximation consists in replacing the 
pyrochlore lattice with a 
tree structure having the same local coordination but lacking
any closed loops beyond those contained in single tetrahedra.
This is illustrated in Fig.~\ref{fig:Husimi tree}.

The Husimi tree of depth $L$ can be seen as being composed of 
$L+1$ 
shells of spins,
which we label by an integer $n$.
We label the outermost shell as $n=0$ and the 
innermost as $n=L$.
The outermost tetrahedra are composed of three spins from shell
$n=0$ and one from shell $n=1$.
Moving inwards, tetrahedra are then composed of three spins from shell
$n=m$ and one from shell $n=m+1$, until 
the central tetrahedron which is composed of four spins from shell
$n=L$.

The partition function, and thermodynamic expectation values of quantities on the tree can be built up by 
successively summing over the states of each shell, working from 
outermost to innermost.
Quantities such as the energy and magnetisation are calculated by
finding their average value on the central tetrahedron, which we
take as being representative of a tetrahedron in the bulk of
the pyrochlore lattice.
This approach has already been shown to be quite successful in
describing spin ice \cite{jaubert08, jaubert-thesis, jaubert14, otsuka18}
and related Ising models \cite{jurcisinova17}.

We once again consider an external field ${\bf h}$
with the largest Cartesian component along the $z$-direction
(i.e. $\cos(\theta) > |\sin(\theta) \sin(\varphi)|, |\sin(\theta) \cos(\varphi)|$.
Other cases can be simply obtained by applying cubic
lattice symmetries to this case.

To somewhat simplify our expressions, we will take the energy of the ground state to be zero, and then reintroduce the original ground state Zeeman energy after the recursion relation has been solved. The original Zeeman energy per spin in the ground state is:
\begin{equation}
    u_0= -\frac{h}{\sqrt{3}}\cos(\theta)
    \label{eq: Ground Zeeman}
\end{equation}

Furthermore, we introduce new Ising variables $\tau_i$
which take the value $+1$ if spin $\vec{S}_i$ has a
positive projection along $\vec{h}$ and $-1$ otherwise.
$\tau_i$ relate to the $\sigma_i$ introduced in Eq. 
(\ref{eq:sigma_introduction}) via:
\begin{eqnarray}
\tau_i = \epsilon_i \sigma_i
\end{eqnarray}
with $\epsilon_i=(1,-1,-1,1)$ respectively for sublattices
$1-4$ (cf. Eq. (\ref{eq: spin directions})).

In calculating the partition function, $Z$, we only include
configurations where the ice rule is obeyed everywhere, i.e.
\begin{eqnarray}
\sum_{i \in t} \sigma_i
=\sum_{i \in t} \epsilon_i \tau_i
=0
\end{eqnarray}
for all tetrahedra, $t$.

The partition function of the Husimi tree is 
\begin{eqnarray}
Z=\sum_{\{ \tau \} }  e^{\sum_j \beta E_j (\tau_j  - 1)/2}
\prod_{t} \left(\delta_{\sum_{i \in t} \epsilon_i \tau_i , 0} \right)
\label{eq:partition_function1}
\end{eqnarray}
where $\sum_{\{ \tau \} }$ is a sum over all configurations of
the Ising variables $\tau_j$, $\prod_{t}$ is a product over
all tetrahedra in the tree, the Kronecker delta $\delta_{\sum_{i \in t} \epsilon_i \tau_i , 0}$ enforces the ice rule on each tetrahedron and $E_j$ is the energy cost of flipping $\tau_j$ against the applied field.

$E_j=2 \epsilon_j \vec{h} \cdot \hat{e}_j$ [cf. Eq. (\ref{eq: spin directions})] depends on which of the four sublattices $j$ belongs to.
$E_j$ can therefore take four possible values which we label
$E_1, E_2, E_3, E_4$ with the subscript now corresponding to the sublattice label.

To make progress with Eq. (\ref{eq:partition_function1}) we
consider it as sum of six terms, corresponding to the six possible arrangements of the central tetrahedron.
Each term in the sum is then a product of the partition function of the four branches, taken with fixed values of the 
spins on layer $n=L$.
This gives us:
\begin{eqnarray}
&&Z=Z_{+1, 1, L} Z_{+1, 2, L} Z_{+1, 3, L}
Z_{+1, 4, L} + \nonumber \\ 
&&
\exp(-\beta (E_1 +E_2))
Z_{-1, 1, L} Z_{-1, 2, L} Z_{+1, 3, L}
Z_{+1, 4, L}+ \nonumber \\
&&
 \exp(-\beta (E_1 +E_3))
Z_{-1, 1, L} Z_{+1, 2, L} Z_{-1, 3, L}
Z_{+1, 4, L} +\nonumber \\
&&
 \exp(-\beta (E_2 +E_4))
Z_{+1, 1, L} Z_{-1, 2, L} Z_{+1, 3, L}
Z_{-1, 4, L}+ \nonumber \\
&&
 \exp(-\beta (E_3 +E_4))
Z_{+1, 1, L} Z_{+1, 2, L} Z_{-1, 3, L}
Z_{-1, 4, L}+
\nonumber \\
&&
 \exp(-\beta (E_1+E_2+E_3 +E_4))
Z_{-1, 1, L} Z_{-1, 2, L} Z_{-1, 3, L}
Z_{-1, 4, L}. \nonumber \\
\label{eq:full_tree_partition_function}
\end{eqnarray}
where $Z_{\tau, i, n}$ is the partition function of a branch terminating on a site of sublattice $i$ at layer $n$, with the value of the terminating spin fixed to $\tau$.

$Z_{\tau, i, n}$ have a recursion relation:
\begin{eqnarray}
&&Z_{\tau, i, n+1}=  \nonumber \\
&&
\left(\prod_{j \neq i} \sum_{\tau_j=\pm1}  \right)
\delta_{\epsilon_i \tau +  \sum_{j \neq i} \epsilon_j \tau_j, 0}
e^{\beta E_j (\tau_j -1)/2} \prod_{j \neq i} Z_{\tau_j, j, n}. \nonumber \\
\label{eq:recursion1}
\end{eqnarray}

To simplify the notation, we define:
\begin{eqnarray}
&&A_n=Z_{+1, 1, n}, \ \ \alpha_n=Z_{-1, 1, n} 
\nonumber \\
&&B_n=Z_{+1, 2, n}, \ \ \beta_n=Z_{-1, 2, n} 
\nonumber \\
&&C_n=Z_{+1, 3, n}, \ \ \gamma_n=Z_{-1, 3, n} 
\nonumber \\
&&D_n=Z_{+1, 4, n}, \ \ \delta_n=Z_{-1, 1, n} 
\label{eq:simplified_partition}
\end{eqnarray}

These can be calculated easily for $n=1$
because this only requires
summing over the configurations of three spins on the outermost layer (see Fig. \ref{fig:Husimi tree}).
For example
\begin{align}
    A_1=&1+e^{-\beta(E_3+E_4)}+e^{-\beta(E_2+E_4)} \label{eq: Husimi single up} \\
    \alpha_{1}=&e^{-\beta(E_2+E_3+E_4)}+e^{-\beta E_2}+e^{-\beta E_3} \label{eq: Husimi single down}.
\end{align}

The partition function of the full tree [Eq. (\ref{eq:full_tree_partition_function})] can then be
obtained by repeatedly applying the recursion relation
Eq. (\ref{eq:recursion1}) to calculate $Z_{\tau, i, L}$.

The sequence thus generated is, however, diverging 
for $L\to\infty$.
Fortunately, useful thermodynamic quantities such as the internal energy and magnetisation can be expressed in terms of four new sequences, which all converge to a finite limit with increasing $L$. These four new sequences are:
\begin{multline}
    Y_n=\frac{\alpha_n}{A_n}e^{-\beta E_1}; \ X_n=\frac{\beta_n}{B_n}e^{-\beta E_2}; \\ W_n=\frac{\gamma_n}{C_n}e^{-\beta E_3}; \ V_n=\frac{\delta_n}{D_n}e^{-\beta E_4}
    \label{eq: Ratio sequences}
\end{multline}

The recursion relations obeyed by these sequences are given in Appendix \ref{app:husimi}.

To extract useful approximations for thermodynamic quantities for spin ice, we make the assumption that the central tetrahedron of the tree is representative of a tetrahedron in the bulk of the pyrochlore lattice, and that its mean magnetisation and internal energy are good approximations for the magnetisation and internal energy of spin ice per `A' tetrahedron. 

Using this approach, we use Eq. (\ref{eq: Ground Zeeman}) to write down an expression for the internal Zeeman energy per spin as:
\begin{eqnarray}
   &&  U = -\frac{h}{\sqrt{3}}\cos(\theta) + \frac{1}{ 4 R_L} \bigg((E_1+E_3) Y_L W_L  \nonumber \\
   && \qquad 
   +(E_1+E_2) Y_L X_L 
   + (E_3+E_4) V_L W_L \nonumber \\
   && \qquad +
    (E_2+E_4) X_L V_L 
   \nonumber
    \\
  && \qquad  
  +(E_1+E_2+E_3+E_4)Y_L X_L W_L V_L \bigg)
    \label{eq: Internal energy}
\end{eqnarray}
where
\begin{eqnarray}
&&R_L=1+Y_L W_L + Y_L X_L+ V_L W_L + 
\nonumber \\
&& \qquad \qquad X_L V_L +Y_L X_L W_L V_L.
\label{eq:RL}
\end{eqnarray}

\subsection{Heat Capacity and Entropy}
\label{subsec:heatcapacity}

To gain some initial insight into the dependence of the thermodynamic quantities on field direction, we consider the
heat capacity, $C(T)$, and entropy, $S(T)$.

\begin{figure}[h]
    \centering
    \subfigure[]{
    \includegraphics[width=7cm]{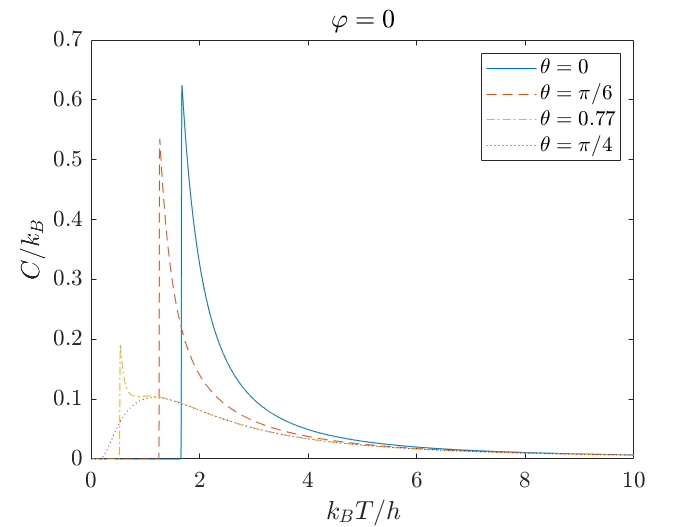}}
    \subfigure[]{
    \includegraphics[width=7cm]{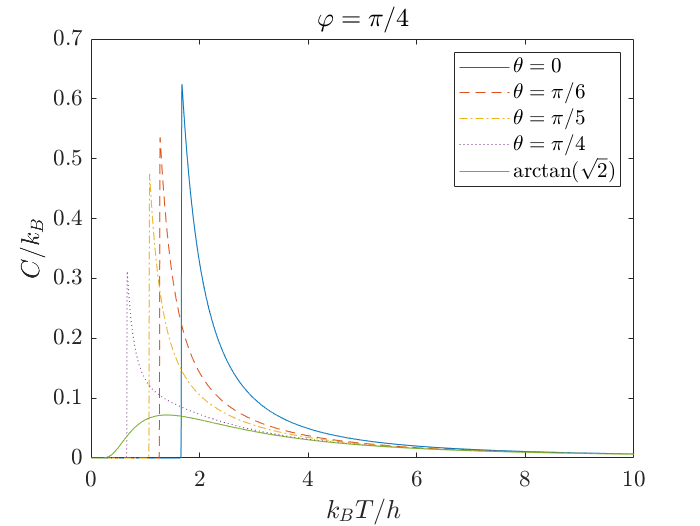}}
    \caption{Heat capacity per spin as a function of temperature $C(T)$, using the Husimi tree approximation,  for various values of $\theta$
    with $\varphi=0$ [(a)] and
    $\varphi=\frac{\pi}{4}$ [(b)]. 1000 shells are used for the Husimi tree. The heat capacity is discontinuous, but not divergent, at the Kasteleyn transition, indicating that there is no latent heat associated with the transition.
    The discontinuity gets progressively smaller and moves to lower temperature
    as the field direction is tuned towards the phase boundaries of Fig. \ref{fig:Magnetic phase diagram}
    before disappearing entirely at the boundaries.}
    \label{fig:Heat capacity against Temperature}
\end{figure}

We calculate the heat capacity in our Husimi tree calculations by calculating the energy per spin 
according to Eq. (\ref{eq: Internal energy}), with $L=1000$ shells, at a series
of temperatures, and then calculating the temperature derivative $C=\left(\frac{\partial U}{\partial T}\right)_{\vec{h}}$ numerically.
The results of this are shown for a series of different
field directions in Fig. \ref{fig:Heat capacity against Temperature}.

Approaching $T_K$ from above, the heat capacity increases sharply, but does not diverge, implying an absence of latent
heat at the transition.
At $T_K$, the heat capacity drops discontinuously to zero, as a result of the complete absence of fluctuations for $T<T_K$.
The value of $T_K$ found in the Husimi tree calculation agrees with the prediction of Eq. (\ref{eq: beta_K}) for all field directions.

Rotating the field direction away from $[001]$ and towards the phase boundaries of Fig. \ref{fig:Magnetic phase diagram}, shifts $T_K$ to lower temperatures and decreases the size of the discontinuity in $C$.
At the phase boundaries, the discontinuity disappears and 
$C$ exhibits only a smooth maximum.

The entropy change between two temperatures $T_1$ and $T_2$
is obtained from the integral of $C/T$:
\begin{equation}
    \Delta S= \int_{T_1}^{T_2} \frac{C(T)}{T} dT.
    \label{eq: Entropy change}
\end{equation}

For field directions for which a Kasteleyn transition occurs
(i.e. those not lying on the phase boundaries of Fig. \ref{fig:Magnetic phase diagram}) we know
that $S(T<T_K)=0$, since all fluctuations are suppressed for
$T<T_K$.
For such field directions we can therefore
obtain the absolute entropy per site by integrating up from $T_K$:
\begin{equation}
    S(T)= \int_{T_K}^{T} \frac{C(T')}{T'} dT'.
    \label{eq: Absolute Entropy}
\end{equation}
For large temperatures we find that this calculation recovers the
Pauling entropy:
\begin{eqnarray}
S_{\rm Pauling}= \frac{1}{2} \log\left( \frac{3}{2} \right)
\approx 0.203
\end{eqnarray}
for all field directions not lying on the phase boundaries.

With this knowledge in hand we can then use the assumption 
that $S(T)$ is independent of field direction for $T\gg h$
to calculate the residual ($T=0$) entropy $S_0$ on the phase boundaries:
\begin{eqnarray}
S_0=S_{\rm Pauling} - \int_{0}^{\infty} \frac{C(T')}{T'} dT'.
\end{eqnarray}

We find that $S_0=0$ along the phase boundaries, apart from
at the kagome ice points, which occur where three phases meet in Fig. \ref{fig:Magnetic phase diagram}.
This is because the lines of phase boundary apart from the kagome ice points have only sub-extensive ground state degeneracy.

At the kagome ice points we find
\begin{eqnarray}
S_0\approx0.072
\end{eqnarray}
in agreement with the modified Pauling estimate \cite{udagawa02} for kagome ice $S^{KI}_{\rm Pauling}=\frac{1}{4}\log\left( \frac{4}{3} \right)$.
This differs by about $10\%$ from the exact result for the entropy of kagome ice \cite{udagawa02, moessner01-ising}. 
The fact that we find agreement with the Pauling approximation, rather than the exact result, is a consequence of using the Husimi tree approximation.

\begin{figure}[h]
    \centering
    \subfigure[]{
    \includegraphics[width=7cm]{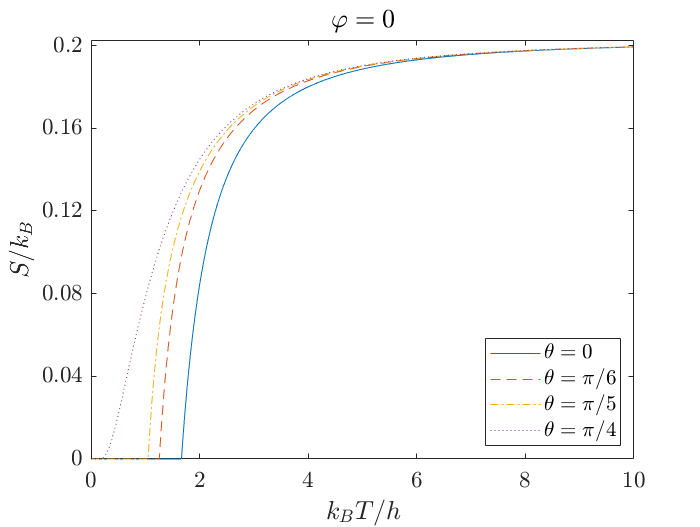}}
    \subfigure[]{
    \includegraphics[width=7cm]{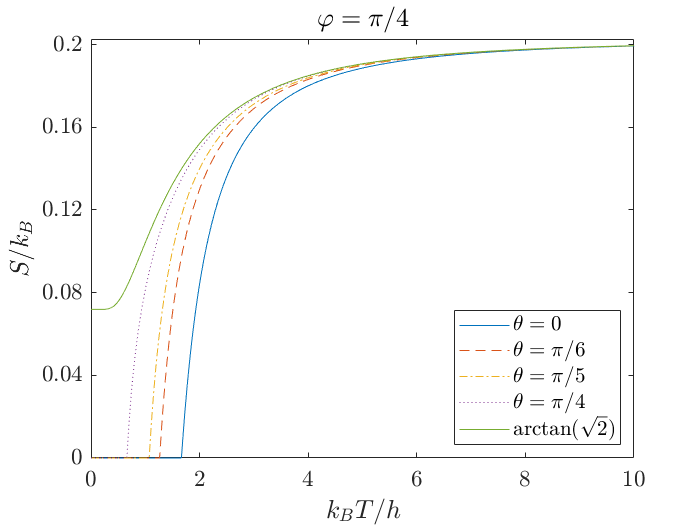}}
    \caption{Entropy per spin as a function of temperature, $S(T)$, using the Husimi tree approximation, for various values of $\theta$
    with $\varphi=0$ [(a)] and
    $\varphi=\frac{\pi}{4}$ [(b)]. 1000 shells are used in the Husimi tree in all cases. The entropy falls to zero at the Kasteleyn temperature, and remains at zero below the transition. 
    For field directions corresponding to the $T=0$
    phase boundaries in Fig. \ref{fig:Magnetic phase diagram} ($\varphi=0, \theta=\frac{\pi}{4}; \varphi=\frac{\pi}{4}, \theta=\arctan(\sqrt{2})$), no Kasteleyn transition is seen. The ground state entropy is however vanishingly small for all these directions except the kagome ice point ($\varphi=\frac{\pi}{4}, \theta=\arctan(\sqrt{2})$), where a limit of $0.072 k_B$ is found.}
    \label{fig: Entropy against Temperature}
\end{figure}

With the $T\to0$ limit of the entropy now determined for all
parameter sets, we can calculate the absolute entropy for all temperatures and field directions. The results of this 
are shown in Fig. \ref{fig: Entropy against Temperature}.

\subsection{Magnetisation and Magnetic Torque}
\label{subsec:magnetisation}

\begin{figure*}
    \centering
    \includegraphics[width=18cm]{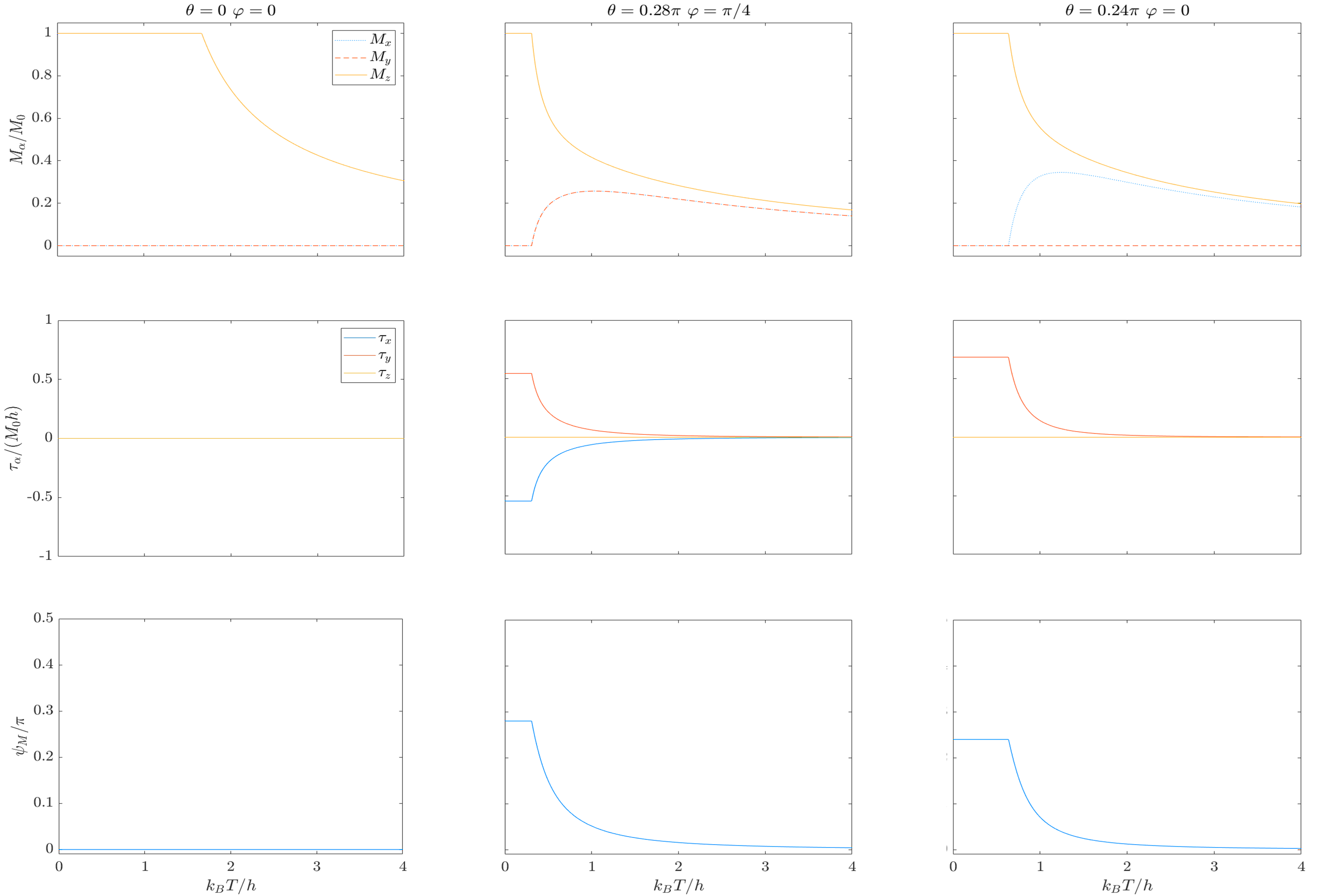}
    \caption{Magnetisation (top row), magnetic torque (middle row) and
    angle between magnetisation and applied field
    (bottom row)
    as a function of temperature for three different field directions.
    The three columns correspond to the three field directions: the [001] direction (left column),
    a slight deviation from the [111] direction (middle)
    and a slight deviation from the [110] direction (right).
    The magnetisation rises rapidly before reaching a plateau at $T=T_K$ for all field directions.
    At high temperatures the magnetisation aligns with the field, as dictated by linear response and the cubic symmetries of the lattice resulting in vanishing torque, $\vec{\tau}$, and angle, $\psi_M$. 
    For fields not aligned with [001], the magnetisation vector rotates continuously away from the field direction as temperature decreases towards $T_K$ resulting in rapid rises in both 
    $\vec{\tau}$ and $\psi_M$.
    }
    \label{fig:mag_and_torque}
\end{figure*}

In this section we present calculations of the magnetisation and magnetic torque as a function of temperature.

Much like the internal energy, we take the mean magnetisation per four spins of spin ice to be the mean magnetisation of the central Husimi tree tetrahedron. 
This gives us the following expressions for the components of the magnetisation, presented as a fraction of the saturation magnetisation $M_0=\frac{N}{\sqrt{3}}$, with $N$ the total number of spins in the system:
\begin{align}
\frac{M_x}{M_0}=&\frac{X_L V_L - Y_L W_L}{R_L} \label{eq: X magnetisation}\\
\frac{M_y}{M_0}=&\frac{V_L W_L - Y_L X_L}{R_L} \label{eq: Y magnetisation}\\
\frac{M_z}{M_0}=&\frac{1-Y_L X_L W_L V_L}{R_L} \label{eq: Z magnetisation}
\end{align}
where $R_L$ is defined by Eq. (\ref{eq:RL}).

The torque acting on the system is found by taking the cross product between the magnetisation and the external field:
\begin{equation}
    \vec{\tau}= \vec{M} \times \vec{h}.
\end{equation}
We present the results for the torque in units
of $M_0 |\vec{h}|$, the value that would be obtained in an extreme limit where the system is polarized in a direction orthogonal to the field.
We also present calculations for the angle between
between the magnetisation and the field:
\begin{eqnarray}
\psi_{M}=\arccos{\left( \frac{\vec{M} \cdot \vec{h}}{|\vec{M}||\vec{h}|} \right)}
\end{eqnarray}

The evolution of $\vec{M}$, $\vec{\tau}$ and $\psi_M$ are
shown as a function of temperature
for three different field directions in Fig. \ref{fig:mag_and_torque}.

At high temperatures $T \gg h$, the magnetisation is determined by the linear response:
\begin{eqnarray}
M_{\alpha} = \sum_{\beta} \chi_{\alpha \beta} h_{\beta}
\end{eqnarray}
and since the cubic symmetry of the lattice requires $\chi_{\alpha\beta}=\chi \delta_{\alpha \beta}$
we have ${\vec M} \parallel \vec{h}$.
$\vec{\tau}$ and $\psi_M$ therefore vanish in the high temperature limit for all field directions, as seen in Fig. \ref{fig:mag_and_torque}.

On the other hand, for $T<T_K$, ${\vec M}$ aligns 
along whichever $\langle001\rangle$ axis makes
the smallest angle with ${\vec h}$.
For generic field directions, this angle may be 
significant, and a large magnetic torque is present in the ordered phase.

The evolution from ${\vec M}\parallel {\vec h}$ to  ${\vec M}\parallel \langle001\rangle$  happens via a continuous rotation of the magnetisation relative to the field as the system is cooled. This process accelerates as $T$ approaches $T_K$ from above.
The central and right columns of 
Fig. \ref{fig:mag_and_torque}
show this for two field directions,
one close to the $[111]$ direction
the other close to the $[101]$ direction.
The value of the magnetic torque obtained as $T \to T_K$ is in both cases a significant fraction of the maximum possible value $M_0 |\vec{h}|$,
illustrating the strength of this effect.

The small misalignment from the 
$[111]$ and $[101]$ assumed in the calculations in Fig. \ref{fig:mag_and_torque} is important.
If the field were exactly aligned 
with the high symmetry direction $\vec{\tau}$ and $\psi_M$ would vanish at all temperatures, by symmetry.
The misalignment allows $\vec{\tau}$ and $\psi_M$ to appear without breaking symmetries, and also makes $T_K$ finite. 
The approach to $T_K$ is then associated with a rapid growth of 
$\vec{\tau}$ and $\psi_M$ as the magnetisation rotates towards a $\langle001\rangle$ direction.

In this sense, the high symmetry alignments along $\langle 111 \rangle$ and $\langle 110 \rangle$
are unstable at low temperatures - small misalignments will produce a torque which makes the misalignment worse.

\subsection{Rotational Magnetocaloric Effect}
\label{subsec:magnetocaloric}

As a final application of our theory, we present calculations of the Rotational Magnetocaloric Effect (RMCE).

As shown in Fig. \ref{fig:Kasteleyn_Temp_plot}, the Kasteleyn temperature $T_K$ depends sensitively on the direction of the applied field. 
The surface $T_K(\theta, \phi)$ can be seen as as surface of constant, vanishing, entropy $S=0$.
This already suggests other constant entropy surfaces, close to $S=0$, will vary strongly with field direction, which in turn implies that an adiabatic (constant entropy) rotation of the crystal relative to the field can induce large temperature changes.
This is the rotational magnetocaloric effect.

\begin{figure}[h]
\centering
\subfigure[ ]{
\includegraphics[width=7cm]{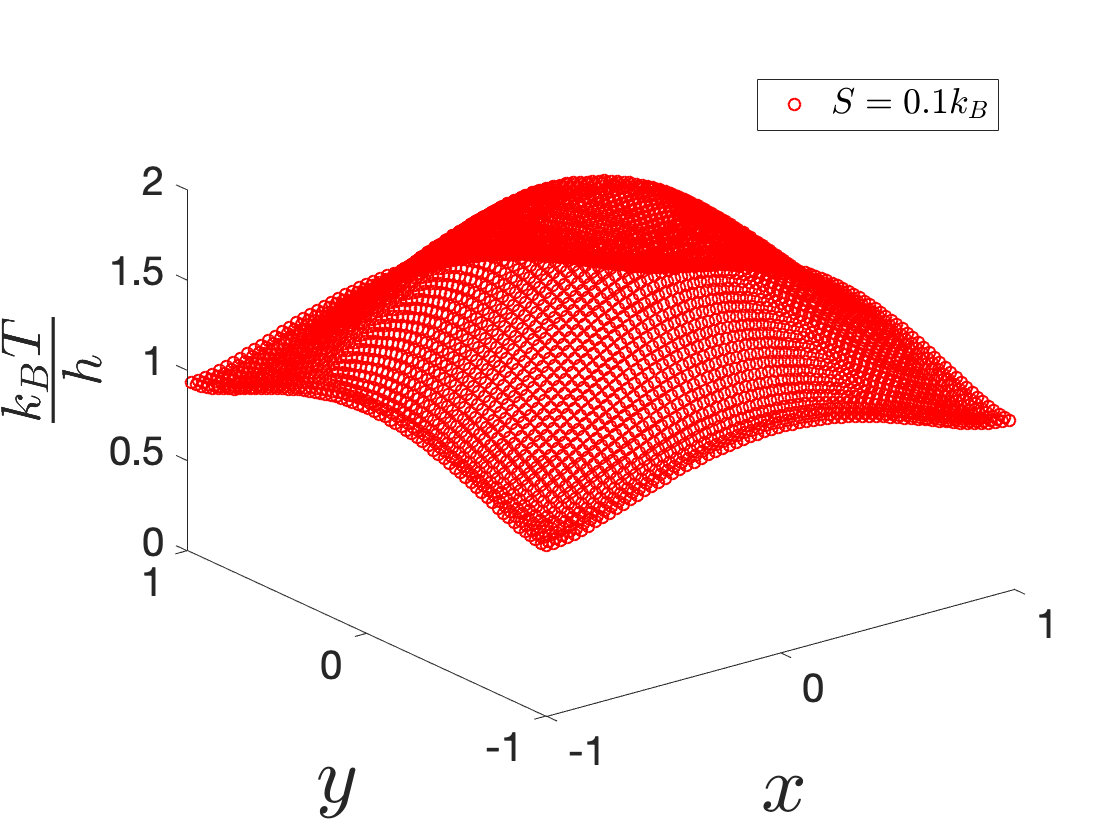}} \\
\subfigure[ ]{
\includegraphics[width=7cm]{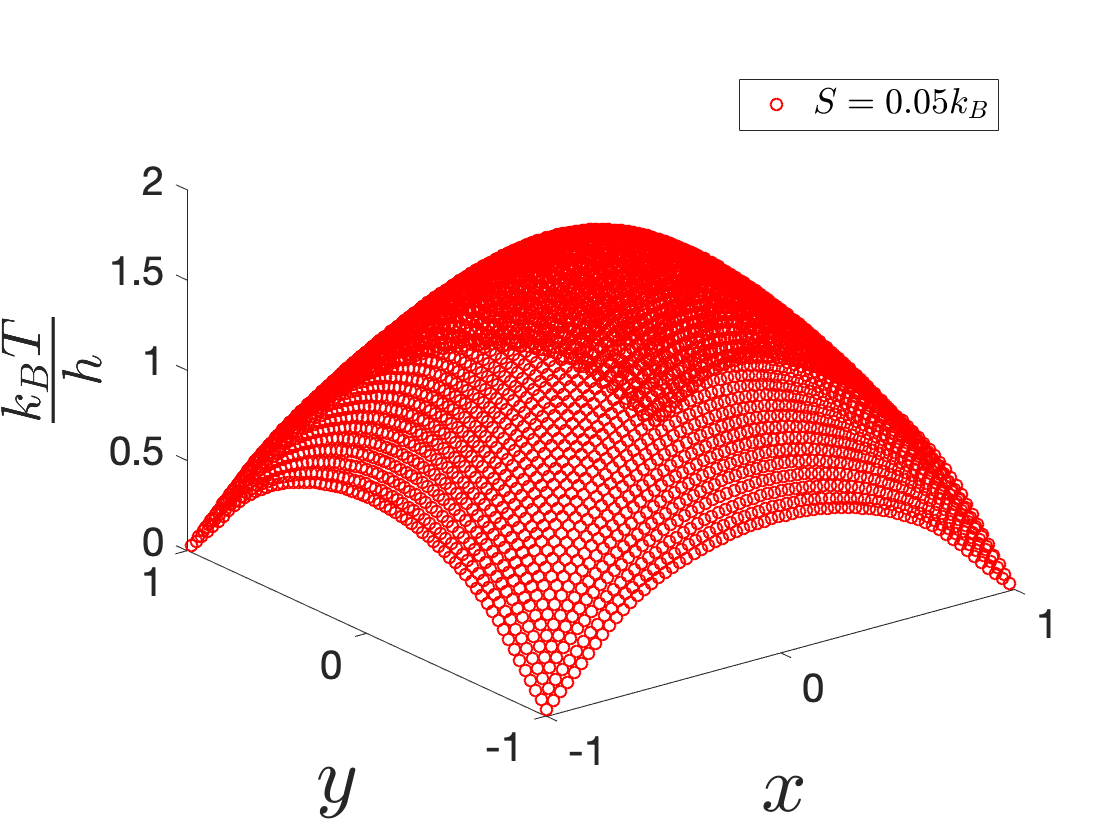}}
\caption{Temperature at constant entropy, as a function of field
direction, for two values of the entropy per spin: $S=0.1k_B$ [(a)]
and $S=0.05k_B$ [(b)].
At $S=0.1 k_B$, adiabatically rotating the field away from the $[001]$ rotation, towards the $[101]$ or $[111]$ directions can reduce the temperature by a factor of $\sim 2$.
For $S=0.05 k_B$, much larger changes are possible, particularly by 
rotating towards the $[111]$ direction.
If $S<S_{KI}$, the entropy of kagome ice, then within the idealized model studied here one
can tune to arbitrarily low temperatures by rotating towards the $[111]$
direction, although this would not be true in a real system in which the degeneracy of kagome ice would not exact.
The relationship between the coordinates $x,y$
and the field direction is given by Eqs. (\ref{eq:h_direction}), (\ref{eq: x parameterisation})-(\ref{eq: theta parameterisation}).
}
\label{fig:T_surfaces}
\end{figure}

\begin{figure*}[t]
\centering
\includegraphics[width=7cm]{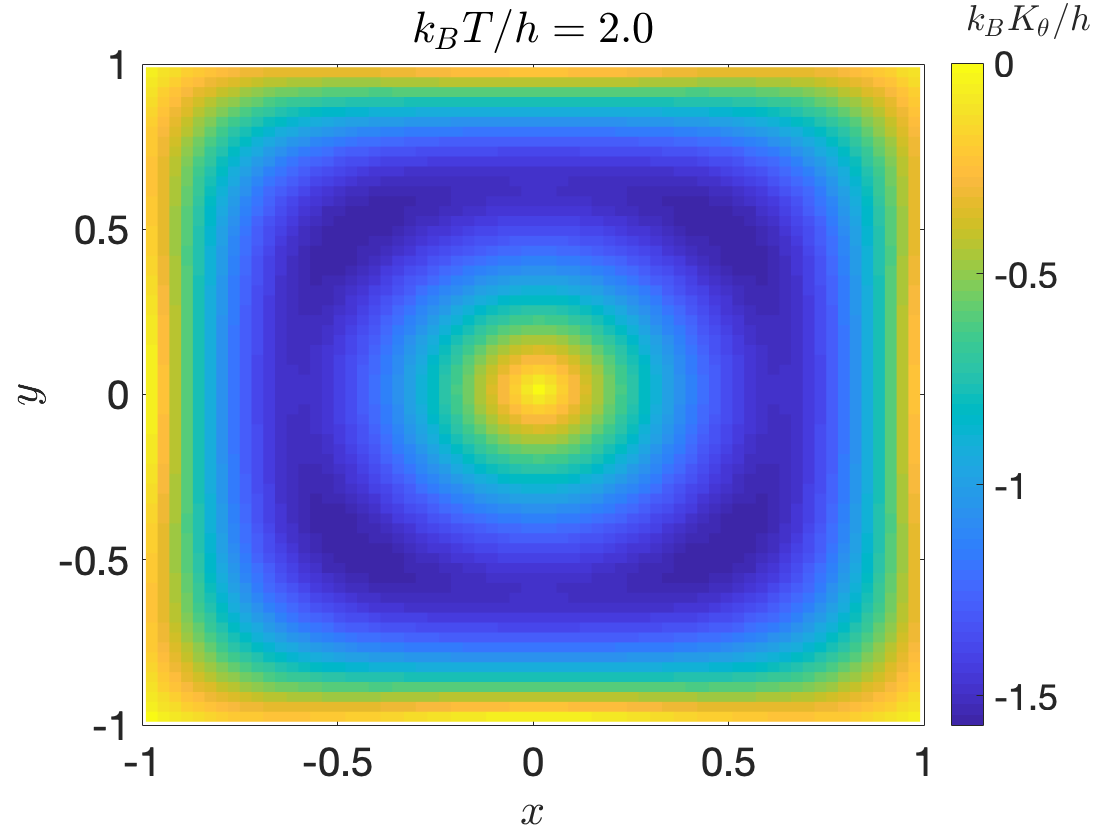} \
\includegraphics[width=7cm]{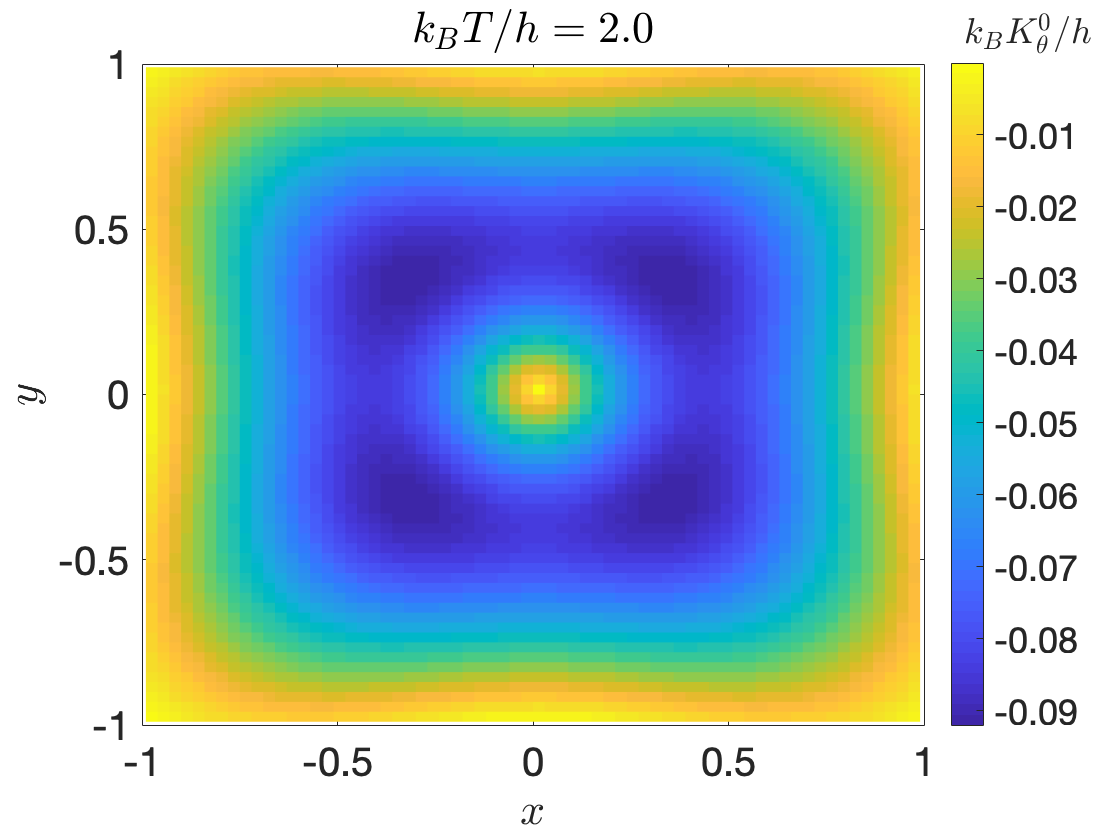} \\
\includegraphics[width=7cm]{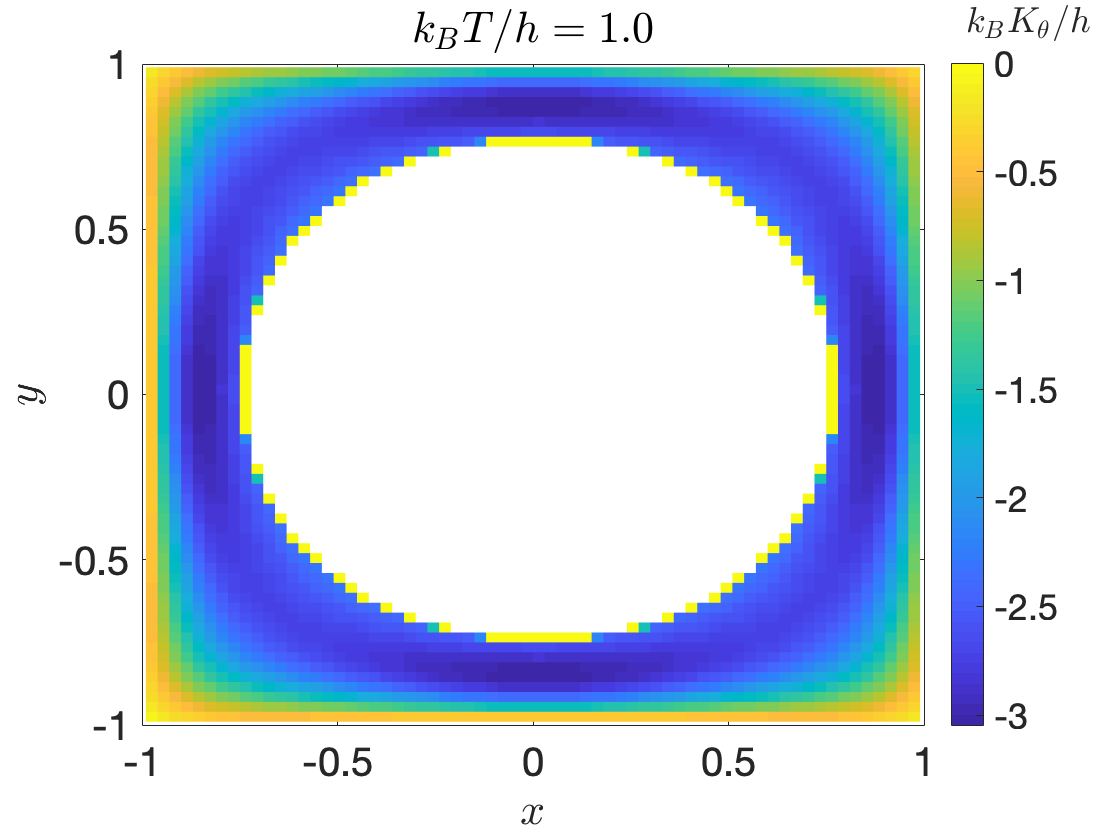} \
\includegraphics[width=7cm]{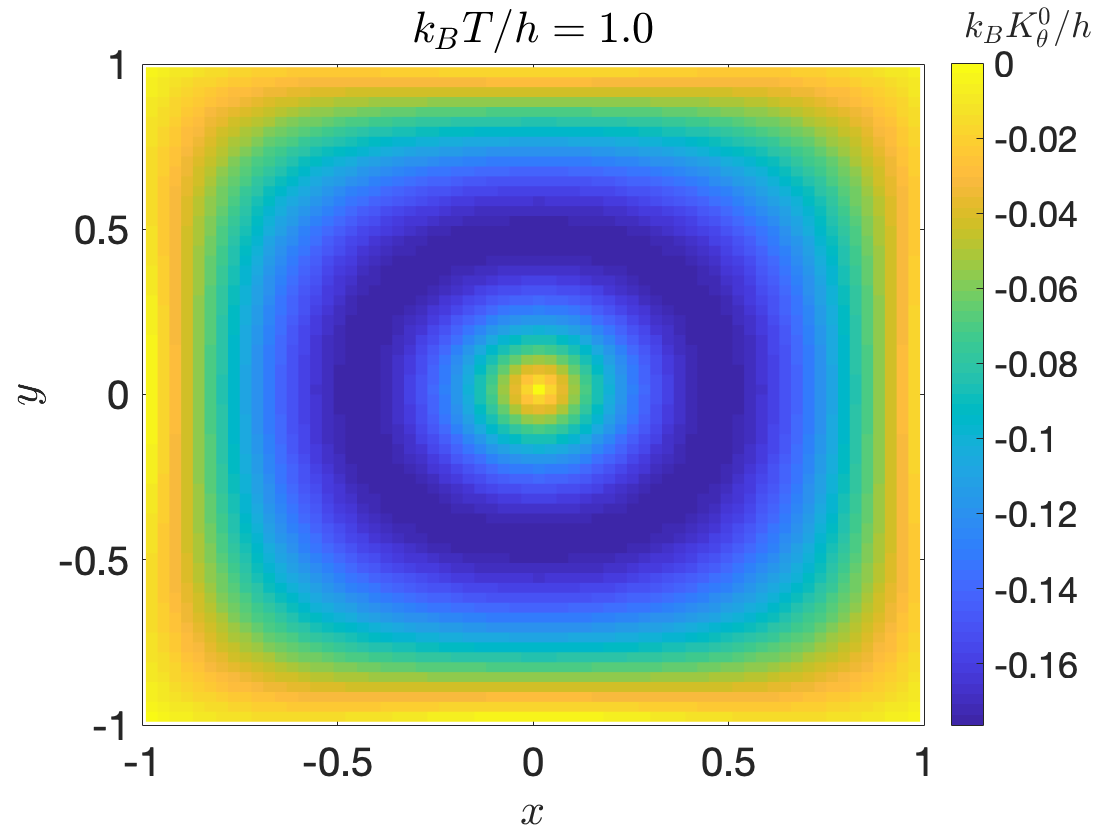} 
\caption{Angular cooling rate $K_{\theta}$ for spin ice (left column) and a system of non-interacting spins on the pyrochlore lattice with the same local anisotropy (right column).
Results are shown for $k_B T = 2h$ (top row) and 
$k_B T = h$ (bottom row).
While the angular variation is similar between the
interacting and non-interacting calculations, the overall magnitude of the cooling rate is an order of magnitude stronger for the interacting case.
This demonstrates that the interactions encoded in the ice rule magnify the RMCE.
The white region in the bottom left panel is in the field induced ordered phase $T<T_K$, in which case there is no RMCE.
The relationship between the coordinates $x,y$
and the field direction is given by Eqs. (\ref{eq:h_direction}), (\ref{eq: x parameterisation})-(\ref{eq: theta parameterisation}).
} 
\label{fig:thetacool}
\end{figure*}

\begin{figure*}[t]
\centering
\includegraphics[width=7cm]{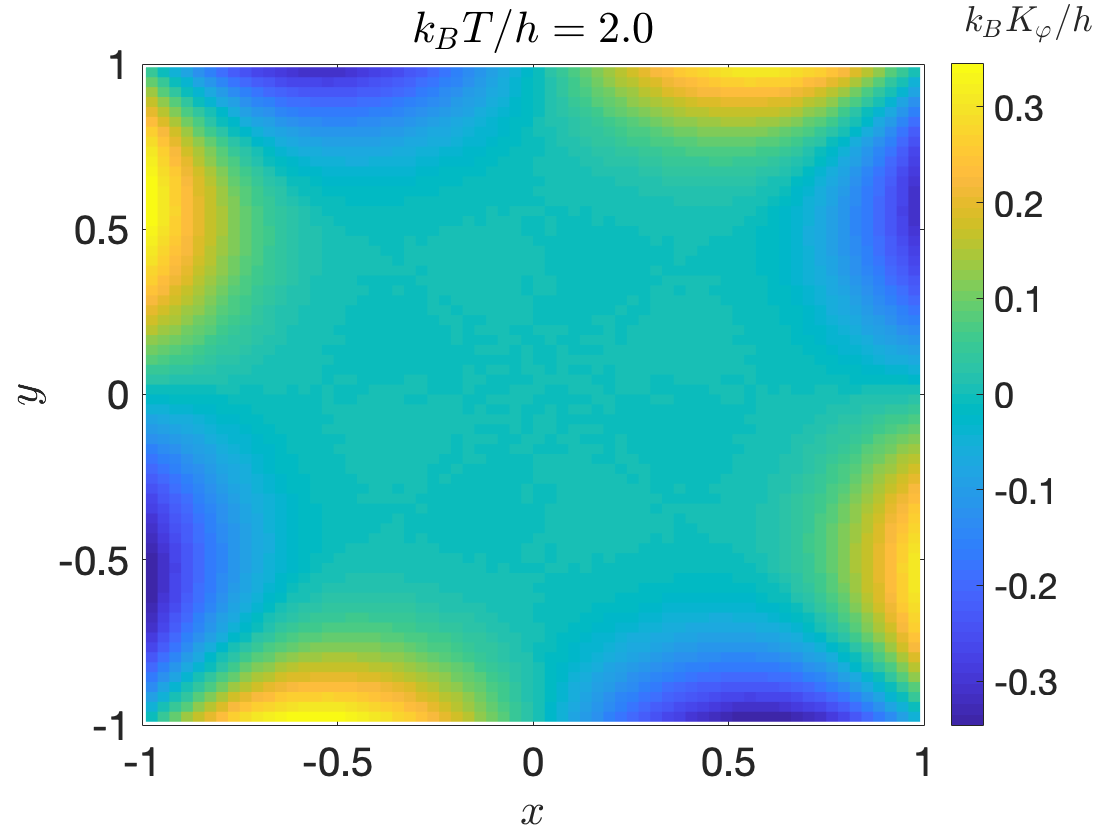} \
\includegraphics[width=7cm]{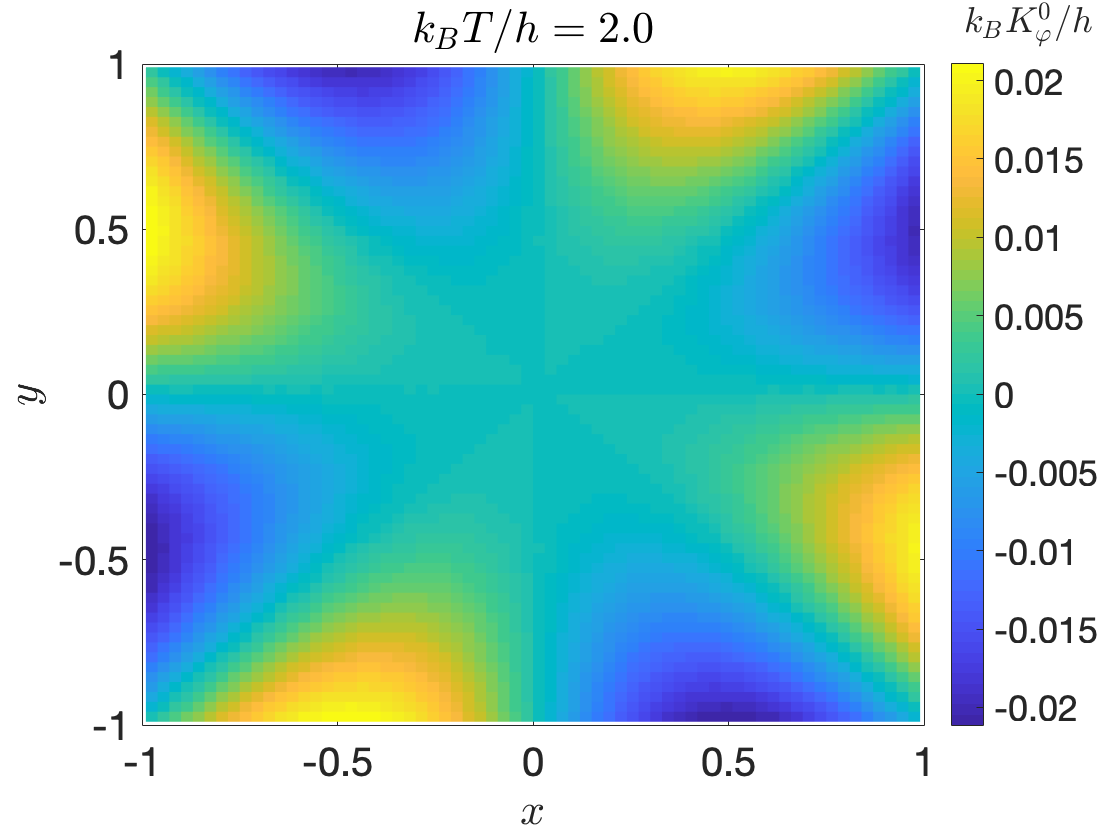} \\
\includegraphics[width=7cm]{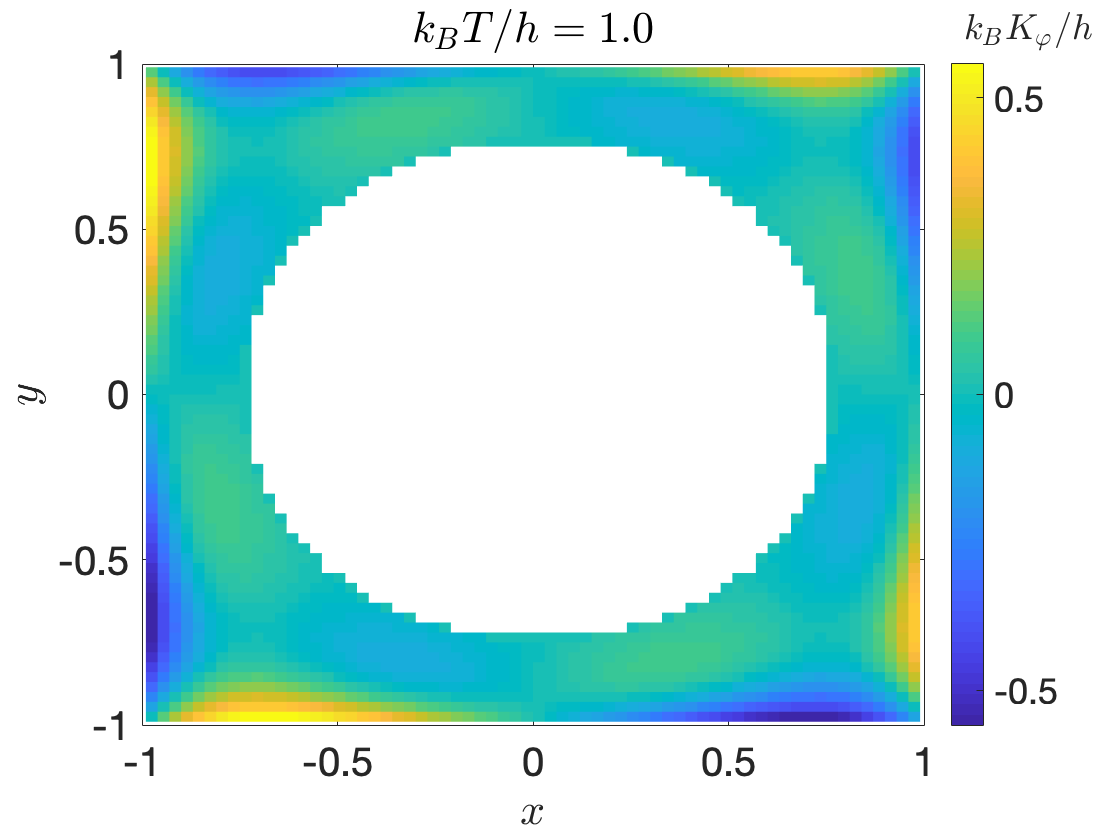} \
\includegraphics[width=7cm]{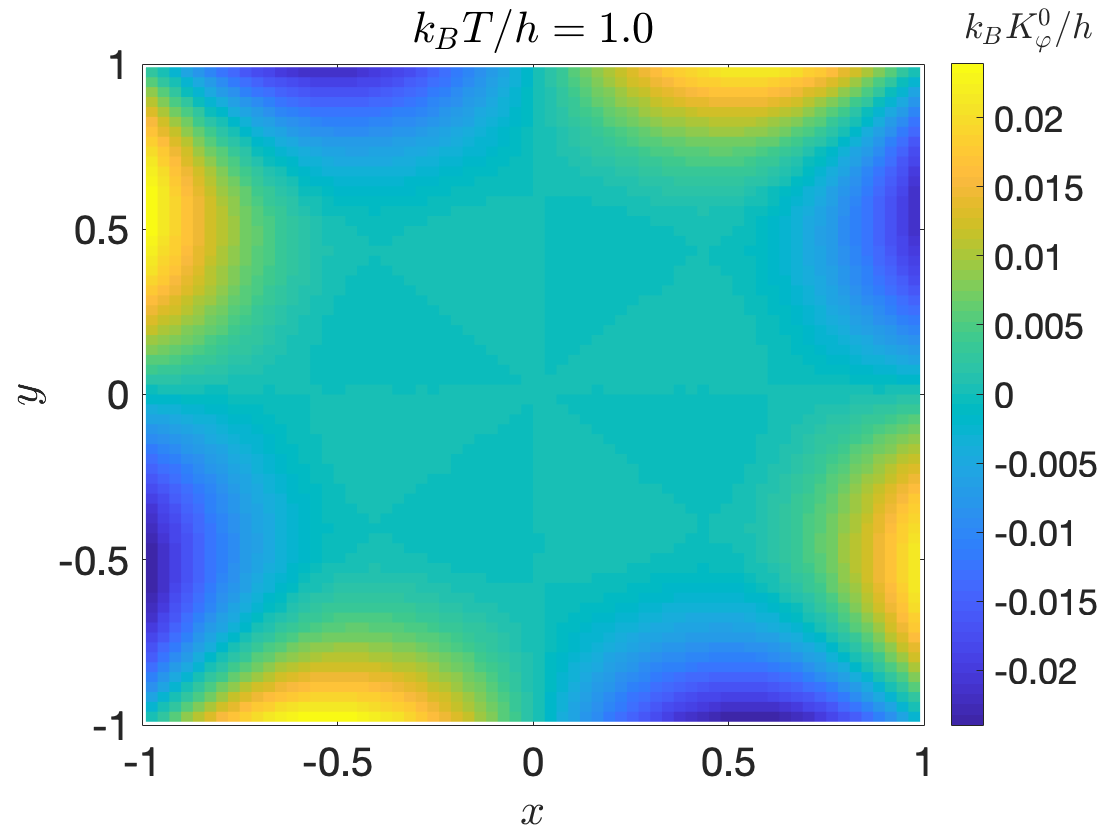} 
\caption{Angular cooling rate $K_{\varphi}$ for spin ice (left column) and a system of non-interacting spins on the pyrochlore lattice with the same local anisotropy (right column).
Results are shown for $k_B T = 2h$ (top row) and 
$k_B T = h$ (bottom row).
The angular variation is similar between the
interacting and non-interacting systems, 
but overall magnitude of the cooling rate is an order of magnitude stronger in the interacting case.
$K_{\varphi}$ vanishes along the lines $x=\pm y, x=0, y=0$, in both interacting and non-interacting calculations which is a consequence of symmetry.
The white region in the bottom left panel is in the field induced ordered phase $T<T_K$, in which case there is no RMCE.
The relationship between the coordinates $x,y$
and the field direction is given by Eqs. (\ref{eq:h_direction}), (\ref{eq: x parameterisation})-(\ref{eq: theta parameterisation}).
}
\label{fig:phicool}
\end{figure*}

The variation of temperature with field direction at constant entropy
is shown for two values entropy in Fig. \ref{fig:T_surfaces}.
Adiabatic rotations which move the field away from the $[001]$ crystal
direction reduce the temperature substantially.
A temperature reduction by a factor of $\sim 2$ can be obtained at $S=0.1k_B$
and an even stronger cooling effect is obtained with lower entropies (lower starting temperature).

Entropies below that of kagome ice $S<S_{KI}$ allow cooling to arbitrarily 
low temperature, at the level of our idealized model, by rotating the field towards the $[111]$ crystal direction.
This would be prevented in a real system by the presence of small perturbations to the Hamiltonian which lift the kagome ice degeneracy.

We define cooling rates for adiabatic rotation of the field direction as follows, with a negative cooling rate indicating a decrease in temperature for increasing angle:
\begin{align}
    K_{\theta}=&\left(\frac{\partial T}{\partial \theta}\right)_{S} = \frac{-T}{C (T)} \left(\frac{\partial S}{\partial \theta}\right)_{T}  \label{eq: Cooling rate theta}\\
    K_{\varphi}=&\left(\frac{\partial T}{\partial \varphi}\right)_{S} = \frac{-T}{C (T)} \left(\frac{\partial S}{\partial \varphi}\right)_{T} \label{eq: Cooling rate phi}
\end{align}
with $C(T)$ being the heat capacity calculated in Section \ref{subsec:heatcapacity}.
These relationships follow from the reciprocal relation for partial derivatives.

To provide a benchmark against which we can compare the
RMCE in the spin ice model, we also calculate the cooling rates
$K_{\theta}^0$ and $K_{\varphi}^0$ of a system of non-interacting spins 
on the pyrochlore lattice with the same local anisotropy.
This benchmark system is described by a Hamiltonian
\begin{eqnarray}
H_0=-\vec{h} \cdot \sum_i \vec{S}_i - D_{SI} \sum_i \left( 
\vec{S}_i \cdot \vec{e}_i \right)^2
\end{eqnarray}
with single-ion anisotropy $D_{SI}\to \infty$.
This corresponds to 
Eq. (\ref{eq: Hamiltonian}) without the exchange interaction term, $J$.
Thus by comparing $K_{\theta}$ and  $K_{\varphi}$ to 
$K_{\theta}^0$ and $K_{\varphi}^0$ we can observe the effect of the
interactions encoded in the ice rule on the RMCE.
The calculation of the non-interacting cooling rates $K_{\theta}^0$ and $K_{\varphi}^0$ is described in Appendix \ref{app:uncoupled_spins}.

For both interacting and non-interacting systems the angular cooling rates decay as $\frac{1}{T}$ at high temperature. 
This follows from Eqs. (\ref{eq: Cooling rate theta}) - (\ref{eq: Cooling rate phi}), and the cubic symmetry of the system which causes the angular derivatives of entropy to vanish as $\frac{1}{T^4}$ at high temperature (see Appendix \ref{app:high_t}).

The angular cooling rates for both interacting and non-interacting systems are shown in Figs. \ref{fig:thetacool}-\ref{fig:phicool}, for
temperatures $k_B T=2 h$ and $k_B T=h$.
From the overall scale of the variation in Figs. \ref{fig:thetacool}-\ref{fig:phicool}
we see that the strong interactions in spin ice enhance the RMCE by roughly an order of magnitude
at these intermediate temperatures.
The azimuthal cooling rate $K_{\varphi}$ varies in sign, vanishing along the lines $\varphi=n \frac{\pi}{4}, n \in \mathbb{Z}$, in both 
interacting and non-interacting calculations.
This is a consequence of the cubic symmetry 
of the lattice.

\section{Summary and Conclusions}
\label{sec:summary}

In this Article we have presented a theory of 
spin ice in the presence of an applied magnetic field with arbitrary orientation.
We have shown that the Kasteleyn transition known for the case of a field oriented along $\langle 001 \rangle$ appears for general field directions, and have calculated the dependence of the Kasteleyn temperature $T_K$ on field direction.
We find that $T_K$ vanishes along certain special lines of field-orientation-space.

In real spin ices, the presence of a finite
density of monopoles - neglected in our calculation turns the Kasteleyn transition from a sharp transition into a crossover.
This crossover temperature can be extracted from magnetisation measurements \cite{pili21} and our 
predictions regarding the behaviour of $T_K$ as a function of field direction --  in particular the singular behavior approaching the phase boundaries of Fig. \ref{fig:Magnetic phase diagram} -- could thus be tested.

We have further investigated the thermodynamics 
of the Coulomb phase for $T>T_K$ using a Husimi tree approximation, with an emphasis on those
properties related to the anisotropic response
of spin ice to a magnetic field.
We find that fields oriented away from high-symmetry directions generate large magnetic torque as $T$ approaches $T_K$ from above.
Moreover, the strong dependence of the entropy
on field direction leads to a rotational magnetocaloric effect by which the system can be cooled or heated using adiabatic rotations of the crystal relative to the applied field.
This effect is enhanced significantly above what
would be expected for a non-interacting system with the same magnetic anisotropy.

Kittaka {\it et al.} have measured the RMCE
in crystals of Dy$_2$Ti$_2$O$_7$ \cite{kittaka18}.
Our results cannot be directly compared with their's, because their measurements
were carried out in a field and temperature regime in which ice-rule violations (monopoles) are important, and these are 
absent from our description.

Our results provide a case study of enhanced RMCE in a frustrated system, and affirm the usefulness of RMCE as a probe
of exotic physics in anisotropic magnets.
It would be interesting to apply these ideas to putative quantum spin ices, particularly those with a multipolar nature such as the Pr- or Ce- based pyrochlores \cite{
onoda10, petit16,
sibille18, gaudet19, gao19, sibille20, smith21, bhardwaj21}, in which RMCE could provide 
an alternative way of constraining the frustrated multipolar interactions.

\section*{Acknowledgments}

This work was carried out as 
part of the internship program
of the Max Planck Institute for
the Physics of Complex Systems.

\appendix
\section{Recursion relations used in Husimi tree calculation}
\label{app:husimi}

In this Appendix we present the recursion relations used in the Husimi tree calculations of Section \ref{sec:thermodynamics}.

The 8 sequences of partition functions in Eq. (\ref{eq:simplified_partition}) obey the following
recursion relations:
\begin{multline}
     A_{n+1}=B_n C_n D_n+B_n \gamma_n \delta_n e^{-\beta(E_3+E_4)} \\ +\beta_n C_n \delta_n e^{-\beta(E_2+E_4)}
    \label{eq: A recursion relation (again)}
\end{multline}
\begin{multline}
     \alpha_{n+1}=\beta_n \gamma_n \delta_n e^{-\beta(E_2+E_3+E_4)}+\beta_n C_n D_n e^{-\beta(E_2)} \\ +B_n \gamma_n D_n e^{-\beta(E_3)}
    \label{eq: Alpha recursion relation}
\end{multline}
\begin{multline}
      B_{n+1}=A_n D_n C_n+\alpha_n D_n \gamma_n e^{-\beta(E_1+E_3)} \\ + A_n \delta_n \gamma_n e^{-\beta(E_4+E_3)}
    \label{eq: B recursion relation}
\end{multline}
\begin{multline}
     \beta_{n+1}=\alpha_n \delta_n \gamma_n e^{-\beta(E_1+E_4+E_3)}+\alpha_n D_n C_n e^{-\beta(E_1)} \\ +A_n \delta_n C_n e^{-\beta(E_4)}
    \label{eq: Beta recursion relation}
\end{multline}
\begin{multline}
     C_{n+1}=D_n A_n B_n+D_n \alpha_n \beta_n e^{-\beta(E_2+E_1)} \\ + \delta_n A_n \beta_n e^{-\beta(E_4+E_2)}
    \label{eq: C recursion relation}
\end{multline}
\begin{multline}
     \gamma_{n+1}=\delta_n \alpha_n \beta_n e^{-\beta(E_4+E_1+E_2)}+D_n \alpha_n B_n e^{-\beta(E_1)} \\ +\delta_n A_n B_n e^{-\beta(E_4)}
    \label{eq: Gamma recursion relation}
\end{multline}
\begin{multline}
    D_{n+1}=C_n B_n A_n+\gamma_n B_n \alpha_n e^{-\beta(E_3+E_1)} \\ + C_n \beta_n \alpha_n e^{-\beta(E_2+E_1)}
    \label{eq: D recursion relation}
\end{multline}
\begin{multline}
     \delta_{n+1}=\gamma_n \beta_n \alpha_n e^{-\beta(E_3+E_2+E_1)}+C_n \beta_n A_n e^{-\beta(E_2)} \\ +\gamma_n B_n A_n e^{-\beta(E_3)}
    \label{eq: Delta recursion relation}
\end{multline}

The series generated by the recursion relations (\ref{eq: A recursion relation (again)})-(\ref{eq: Delta recursion relation}) does not converge as $L \to  \infty$. However, physical quantities can be written in terms of the variables $Y_n$,$X_n$, $W_n$, $V_n$ for which the corresponding series do
converge.
The recursion relations for these variables are:
\begin{equation}
    Y_{n+1}=e^{-\beta E_1} \frac{X_n W_n V_n +X_n+W_n}{1+V_n W_n + X_n V_n}; \ Y_0=e^{-\beta E_1}
    \label{eq: Y recursion relation (again)}
\end{equation}
\begin{equation}
     X_{n+1}=e^{-\beta E_2} \frac{W_n Y_n V_n +Y_n+V_n}{1+Y_n W_n + V_n W_n}; \ V_0=e^{-\beta E_2}
    \label{eq: X recursion relation}
\end{equation}
\begin{equation}
    W_{n+1}=e^{-\beta E_3} \frac{V_n Y_n X_n +Y_n+V_n}{1+Y_n X_n + X_n V_n}; \ W_0=e^{-\beta E_3}
    \label{eq: W recursion relation}
\end{equation}
\begin{equation}
    V_{n+1}=e^{-\beta E_4} \frac{W_n X_n Y_n +X_n+W_n}{1+Y_n W_n + Y_n X_n}; \ V_0=e^{-\beta E_4}
    \label{eq: V recursion relation}
\end{equation}
\medskip

\section{Details of uncoupled spins calculation}
\label{app:uncoupled_spins}

We here present details of how the cooling rates $C_{r,\theta}^0$ and $C_{r,\varphi}^0$, pertaining to a system of uncoupled spins with the same local anisotropy as spin ice, were calculated. In this model, the ice rule need not be obeyed, and we consider only the single-ion anisotropy and Zeeman terms of Eq. (\ref{eq: Hamiltonian}).
We continue to assume that $D_{SI} \gg |\vec{h}|$ such the spins remain Ising-like and oriented along their local $\langle 111 \rangle$ axis.
Since the model is now non-interacting, the total entropy
can be written as a sum of single-site entropies $S=\sum_i S_i$.

Starting with the ground state configuration of the interacting spin-ice model for an applied [001]
field, we define energies $\Delta_i$ as the difference in Zeeman energy between the two possible orientations of the spin:
\begin{eqnarray}
\Delta_i = 2 \vec{h} \cdot \vec{e}_i
\label{eq:delta_def}
\end{eqnarray}
with the local axes $\vec{e}_i$ defined for each sublattice
in Eq. (\ref{eq: spin directions}).

For each sublattice $i$, we can write down a single spin partition function as:
\begin{equation}
    Z_i=e^{-\frac{\beta \Delta_i}{2}}+e^{\frac{\beta \Delta_i}{2}}=2\cosh\left(\frac{\beta \Delta_i}{2}\right)
    \label{eq: Uncoupled single partition function}
\end{equation}

The internal energy can be calculated straightforwardly as:
\begin{equation}
    U_i=-\frac{\partial \log(Z_i)}{\partial \beta}=-\frac{\Delta_i}{2} \tanh\left(\frac{\beta \Delta_i}{2}\right)
    \label{eq: Uncoupled single internal energy}
\end{equation}
and the single spin entropy is
\begin{equation}
    S_i=\log\left(2\cosh\left(\frac{\beta \Delta_i}{2}\right)\right)+\beta U_i
    \label{eq: Uncoupled single entropy}
\end{equation}

To calculate the cooling rate, we need the heat capacity, which we find by differentiating Eq. (\ref{eq: Uncoupled single internal energy}) with respect to $T$ at constant field:
\begin{equation}
    C_i=\frac{\partial U}{\partial T}=\left(\frac{\beta \Delta_i}{2}\right)^2 \sech^2\left(\frac{\beta \Delta_i}{2}\right)
\end{equation}

The total heat capacity and entropy per spin are then 
$C=\frac{1}{4} \sum_{i=1}^4 C_i; S_i =\frac{1}{4}\sum_{i=1}^4 S_i$ with the sums running over the four sublattices.

The rate of entropy change per site with respect to angle is then:
\begin{eqnarray}
\frac{\partial S}{\partial \theta}=\frac{1}{4} \sum_{i=1}^4 \frac{\partial \Delta_i}{\partial \theta}
\frac{\partial S_i}{\partial \Delta_i} 
\label{eq:dSdtheta}
\\
\frac{\partial S}{\partial \phi}=\frac{1}{4} \sum_{i=1}^{4} \frac{\partial \Delta_i}{\partial \phi}
\frac{\partial S_i}{\partial \Delta_i} 
\label{eq:dSdphi}
\end{eqnarray}
Where the derivatives of $\Delta_i$ with respect to angle
follow from Eqs. (\ref{eq:delta_def}) and (\ref{eq:h_direction}):
\begin{eqnarray}
&&\frac{\partial \Delta_i}{\partial \theta} =
2 h \left(
\cos(\theta) \cos(\phi) \vec{e}_i^{\ x}
+
\cos(\theta) \sin(\phi) \vec{e}_i^{\ y}
-
\sin(\theta) \vec{e}_i^{\ z}
\right) \nonumber \\
\\
&&\frac{\partial \Delta_i}{\partial \phi} =
2 h \left(
-\sin(\theta) \sin(\phi) \vec{e}_i^{\ x}
+
\sin(\theta) \cos(\phi) \vec{e}_i^{\ y}
\right) \nonumber 
\\
\end{eqnarray}
and
\begin{eqnarray}
\frac{\partial S_i}{\partial \Delta_i}=
-\frac{1}{4} \beta^2 \Delta_i \sech\left(
\frac{\beta \Delta_i}{2}
\right)^2.
\label{eq:dsddelta}
\end{eqnarray}

\section{High temperature limit of angular
derivatives of entropy}
\label{app:high_t}

Here we showing that the high temperature 
limit of the angular derivatives of entropy
$\frac{\partial S}{\partial \phi}$, $\frac{\partial S}{\partial \theta}$
behave as $\frac{1}{T^4}$ in the high temperature limit.
This follows from the cubic symmetry of the problem.

We start by writing down a series expansion for $S(\beta)$, with $\beta=1/T$, around
$\beta=0$
\begin{eqnarray}
S(\beta)=\sum_{n=0}^{\infty} s_n \beta^n.
\label{eq:s_expansion}
\end{eqnarray}

The magnetic field provides the only energy scale
in the problem for both the case of spin ice
and the non-interacting paramagnet (since we take, $J, D_{SI} \to \infty$.
This means that
$S(\beta)$ must be invariant under the rescaling $\beta \to \beta/\kappa; h\to \kappa h$, and so the coefficients $s_n$ must each scale as $h^n$ with $h$ being the magnitude of $\vec{h}=(h_x, h_y, h_z)$.

Furthermore, the cubic symmetry of the system implies that each coefficient must be invariant under the action of the symmetry group $O_h$ applied to $\vec{h}$.
Thus, the only symmetry allowed forms for the first few terms of Eq. (\ref{eq:s_expansion}) are
\begin{eqnarray}
&& s_0=a; \ 
s_1=0; \ 
s_2=b(h_x^2+h_y^2+h_z^2); \
s_3=0 \nonumber \\
&& s_4=c(h_x^2+h_y^2+h_z^2)^2+d(h_x^4+h_y^4+h_z^4)
\label{eq:scoeffs}
\end{eqnarray}
with $a,b,c,d$ being field and temperature independent constants.
Of the coefficients in Eq. (\ref{eq:scoeffs}), only $s_4$ depends on the field orientation and so the leading $\beta$ dependence of $\frac{\partial S}{\partial \phi}$, $\frac{\partial S}{\partial \theta}$
must be $\propto  \beta^4$.

This argument holds both for the calculations for spin ice in the main text and for the calculations for non-interacting spins on the pyrochlore lattice in Appendix \ref{app:uncoupled_spins}.

It can readily be verified that if we expand Eq. (\ref{eq:dsddelta}) for small $\beta$ and insert the expansion into Eqs. (\ref{eq:dSdtheta})-(\ref{eq:dSdphi}), the 
$\beta^2$ terms cancel and the leading 
$\beta$ dependence is indeed $\sim \beta^4$.

Inserting this dependence into Eqs. (\ref{eq: Cooling rate theta})-(\ref{eq: Cooling rate phi}) and using the fact that heat capacity $C\sim\frac{1}{T^2}$
at high temperature, we find that the angular cooling rates behave as $\sim \frac{1}{T}$.

\bibliography{spinice.bib}
\end{document}